\documentclass[useAMS,usenatbib,usegraphicx]{mn2e}

\usepackage{color}

\title[Ultraviolet Fe~{\sc II} Emission in Fainter Quasars]{Ultraviolet
Fe~{\sc II} Emission in Fainter Quasars: Luminosity Dependences, and the
Influence of Environments}

\author[R.G. Clowes et al.]
{Roger G.~Clowes,$^1$\thanks{E-mail: rgclowes@uclan.ac.uk}
Lutz Haberzettl,$^2$
Srinivasan Raghunathan,$^3$
\newauthor
Gerard M. Williger,$^2$
Sophia M. Mitchell,$^{2,4}$
Ilona K. S\"ochting,$^5$
\newauthor
Matthew J. Graham$^6$
and
Luis E. Campusano$^3$ \\
$^1$ Jeremiah Horrocks Institute, University of Central Lancashire,
Preston PR1 2HE, UK \\
$^2$ Department of Physics and Astronomy, University of Louisville,
Louisville, KY 40292, USA \\
$^3$ Observatorio Astron\'omico Cerro Cal\'an, Departamento de Astronom\'{\i}a,
Universidad de Chile, Casilla 36-D, Santiago, Chile \\
$^4$ Department of Aerospace Engineering ACCEND, University of Cincinnati,
Cincinnati, OH 45221, USA \\
$^5$ Astrophysics, Denys Wilkinson Building, Keble Road,
University of Oxford, Oxford OX1 3RH, UK \\
$^6$ California Institute of Technology, 1200 East California Boulevard,
Pasadena, CA 91125, USA
}

\begin{document}

\date{Accepted 2016 April 28. Received 2016 April 27; in original form
2015 August 31}

\pagerange{\pageref{firstpage}--\pageref{lastpage}} \pubyear{2011}

\maketitle

\label{firstpage}

\begin{abstract}

We investigate the strength of ultraviolet Fe~{\sc II} emission in fainter
quasars compared with brighter quasars for $1.0 \le z \le 1.8$, using the
SDSS (Sloan Digital Sky Survey) DR7QSO catalogue and spectra of Schneider et
al., and the SFQS (SDSS Faint Quasar Survey) catalogue and spectra of Jiang
et al. We quantify the strength of the UV Fe~{\sc II} emission using the
$W2400$ equivalent width of Weymann et al., which is defined between two
rest-frame continuum windows at 2240--2255 and 2665--2695~\AA. The main
results are the following.  ~(1) We find that for $W2400 \ga 25$~\AA\ there
is a {\it universal\/} (i.e.\ for quasars in general) strengthening of
$W2400$ with decreasing intrinsic luminosity, $L3000$.  ~(2) In conjunction
with previous work by Clowes et al., we find that there is a further, {\it
  differential,} strengthening of $W2400$ with decreasing $L3000$ for those
quasars that are members of Large Quasar Groups (LQGs).  ~(3) We find that
increasingly strong $W2400$ tends to be associated with decreasing FWHM of
the neighbouring Mg~{\sc II} $\lambda 2798$ broad emission line.  ~(4) We
suggest that the dependence of $W2400$ on $L3000$ arises from Ly$\alpha$
fluorescence.  ~(5) We find that stronger $W2400$ tends to be associated with
smaller virial estimates from Shen et al.\ of the mass of the central black
hole, by a factor $\sim 2$ between the ultrastrong emitters and the
weak. Stronger $W2400$ emission would correspond to smaller black holes that
are still growing.
The differential effect for LQG members might then arise from
preferentially younger quasars in the LQG environments.

\end{abstract}

\begin{keywords}
galaxies: active -- quasars: emission lines -- quasars: supermassive black
holes -- galaxies: clusters: general -- large-scale structure of Universe.
\end{keywords}

\section{Introduction} \label{secIntroduction}

We investigate the strength of ultraviolet Fe~{\sc II} emission for fainter
quasars compared with brighter quasars for the redshift interval $1.0 \le z
\le 1.8$. We consider both quasars in general and the subset that are members
of Large Quasar Groups (LQGs).  We use the DR7QSO catalogue and spectra
\citep{Schneider2010} of the Sloan Digital Sky Survey (SDSS) to make
comparisons of fainter and brighter quasars within the limit $i \le 19.1$ of
the low-redshift strand of quasar selection \citep{Richards2006,
  Vanden-Berk2005} for the SDSS. We extend the faint limit of the comparisons
to $g \la 21.0$ using the SDSS Faint Quasar Survey (SFQS) of
\citet{Jiang2006}, which actually reaches $g \sim 22.5$. The use of both $g$
and $i$\, in what follows arises from the specifications of these published
quasar samples.

The investigation is a continuation of the work presented in
\citet[][hereafter Clo13b]{Clowes2013b}. That paper found, for a large sample
of LQG members ($i \le 19.1$, $1.1 \le \bar{z}_{LQG} \le 1.5$), a shift in
the $W2400$ equivalent width \citep{Weymann1991} of $\sim 1$~\AA\ ($0.97 \pm
0.33$~\AA) compared with a matched control sample of non-members. It also
found a tentative indication that the shift in $W2400$ increased to fainter
$i$ magnitudes ($1.31 \pm 0.36$~\AA\ for $18.0 \le i \le 19.1$). Furthermore,
the strongest $W2400$ emitters within the LQGs had preferred
nearest-neighbour separations of $\sim$~30--50~Mpc (present epoch) to the
adjacent quasar of any $W2400$ strength, with no such effect being seen for
quasars outside LQGs.

The further investigation of the dependences of the UV Fe~{\sc II} on
intrinsic luminosity and on environment (LQG or non-LQG) has some potential
to improve our understanding both of the causes of strong and ultrastrong UV
Fe~{\sc II} emission in quasars and of the nature of LQGs.

In this paper, we use the DR7QSO catalogue and spectra \citep{Schneider2010}
and the SFQS catalogue and spectra \citep{Jiang2006} to investigate the UV
Fe~{\sc II} properties of fainter quasars relative to brighter quasars. We
also make some use of our own, rather smaller, data sets. We find a universal
dependence (i.e.\ for quasars in general) of the $W2400$ equivalent width on
the intrinsic continuum luminosity, such that $W2400$ is greater for fainter
quasars. We find a further, differential, dependence, in the same sense, for
quasars that are members of LQGs. We avoid calling either of these effects
``a Baldwin effect'' \citep{Baldwin1977} because there appear to be
differences compared with the standard perception of the Baldwin effect
\citep[e.g.][]{Baldwin1977, Zamorani1992, Bian2012}. We find also that
$W2400$ increases as the FWHM of the Mg~{\sc II} $\lambda 2798$ broad
emission line decreases. We suggest an explanation for these universal and
differential dependences of $W2400$ on luminosity, and consider the
implications for the possible nature of LQGs.

The concordance model is adopted for cosmological calculations, with
$\Omega_T = 1$, $\Omega_M = 0.27$, $\Omega_\Lambda = 0.73$, and $H_0 =
70$~kms$^{-1}$Mpc$^{-1}$. All sizes given are proper sizes at the present
epoch. Later in the paper we incorporate estimates of black hole masses from
\citet{Shen2011}; their adopted cosmological parameters differed a little,
with $\Omega_M = 0.3$ and $\Omega_\Lambda = 0.7$.

\subsection{Brief details on large quasar groups} \label{subsecLQGs}

For a comprehensive discussion of LQGs, see \citet{Clowes2012} and
\citet{Clowes2013a}, together with the earlier references given in those
papers. Essentially, LQGs are large structures --- that is, large-scale
overdensities, usually at some specified amplitude --- of quasars, seen in
the early universe. They have sizes $\sim$~70--500 Mpc and memberships of
$\sim$~5--70 quasars.

Our catalogue of LQGs (Clowes, in preparation) was obtained from DR7QSO
quasars with $i \le 19.1$ and $1.0 \le z \le 1.8$. It was used by
\citet{Clowes2012}, \citet{Clowes2013a}, and Clo13b, and descriptions of the
selection of the LQGs can be found in those papers.

\subsection{UV Fe~{\sc II} emission in quasars} \label{subsecFeII}

The UV Fe~{\sc II} problem in quasars and Seyfert galaxies has been known for
more than thirty years --- see \citet{Wills1980} and \citet{Netzer1980}, and
references given in those papers, for some initial discussions of the
observational and theoretical aspects. Essentially, the problem is that the
UV Fe~{\sc II} emission can vary greatly in strength from one quasar to
another, and the variation appears not to simply correspond to iron
abundance. Of particular interest are the rare ``ultrastrong emitters'', a
good example of which is the quasar 2226$-$3905 \citep*{Graham1996}. They
represent only $\sim$~6.6 per cent of all quasars in the redshift range $1.0
\le z \le 1.8$ and with $i \le 19.1$ (Clo13b). Also of particular interest is
the ratio of the UV Fe~{\sc II} flux to that of the nearby Mg~{\sc
  II}~$\lambda$2798 doublet, Fe~{\sc II}(UV)/Mg~{\sc II}, because of its
potential to allow deduction of the abundance ratio Fe/$\alpha$, where
$\alpha$ refers to the $\alpha$-elements O,~Ne,~Mg,~etc. The Fe is generally
thought to be produced on timescales $\sim$~1~Gyr by SNe~Ia and the
$\alpha$-elements on shorter timescales by SNe~II \citep[e.g.][]{Hamann1999,
  Hamann2004}. Measurement of the abundance ratio in quasars could then
conceivably allow deductions about the time of major star formation relative
to the time of quasar activity --- an ``iron-clock''. However, aside from the
question of the extent to which the ratio Fe~{\sc II}(UV)/Mg~{\sc II} relates
to the abundance ratio (see below), the observations show that it is
essentially constant for redshifts $z \la 6.5$ and that star formation
therefore appears to precede quasar activity by perhaps 0.3~Gyr
\citep[e.g.][]{Dietrich2003, Simon2010, DeRosa2011}.

Note, however, that \citet{Reimers2005} present an abundance analysis for a
particular (if unusual) quasar that does not support this notion of an
iron-clock. Furthermore, calculations by \citet{Matteucci2001} show that the
timescale for the maximum rate of SNe~Ia, and hence for the maximum
rate of enrichment,
depends strongly on the adopted conditions (stellar lifetimes, initial mass
function, star-formation rate), being $\sim$~40--50~Myr for an instantaneous
starburst, $\sim$~0.3~Gyr for a typical elliptical galaxy, and
$\sim$~4--5~Gyr for the disk of a Milky-Way-type galaxy. An instantaneous
starburst seems plausible for the central activity of a galaxy that precedes
the quasar activity. \citet{Matteucci2001} emphasise that the commonly-used
timescale of $\sim$~1~Gyr is the timescale at which the production of iron
begins to become important for the solar neighbourhood: it is not the time at
which SNe~Ia begin to occur. Note that the progenitors of SNe~Ia have not yet
been established observationally \citep*{Nomoto2013}, although observations
can constrain the possibilities for models and abundance yields.

Also, \citet{Verner2009} found evidence that Fe~{\sc II}(UV)/Mg~{\sc II}, for
$0.75 < z < 2.2$, increases across the interval $z \sim$~1.8--2.2, relative
to its value at lower redshifts. They interpret this result mainly as a
dependence on intrinsic luminosity (ionising flux) rather than on abundance.

Accounting for the strength of the UV Fe~{\sc II} emission and its variation
between quasars is a complicated problem, with many contributing factors such
as iron abundance, hydrogen density, hydrogen column density, temperature,
ionisation flux (excitation flux, continuum shape), microturbulence,
Ly$\alpha$ fluorescence, and spatial distribution of the emitting gas
\citep[e.g.][]{Verner2003, Verner2004, Leighly2006, Bruhweiler2008,
  Gaskell2009, Kollatschny2011, Kollatschny2013}. The quasars with
ultrastrong emission, being extreme, may be particularly useful for
clarifying the relative importance of different factors. The current view is
that probably iron abundance, Ly$\alpha$ fluorescence, and microturbulence
are all contributing significantly to the ultrastrong emission.

The most detailed modelling of Fe~{\sc II} emission is by \citet{Verner2003},
\citet{Verner2004}, and \citet{Bruhweiler2008}. They consider 830 energy
levels of the Fe$^+$ ion, corresponding to 344035 transitions.
\citet{Verner2004} conclude that iron abundance is not the only factor that
can lead to strong emission and, moreover, that it is not even likely to be
the dominant factor. \citet{Verner2003} discuss the relative importance of
abundance and microturbulence, showing that increasing the iron abundance
from solar to five times solar increases the flux ratio Fe~{\sc
  II}(UV)/Mg~{\sc II} by less than a factor of two. Conversely, increasing
the microturbulence from 5~km~s$^{-1}$ to 25~km~s$^{-1}$ increases the ratio
Fe~{\sc II}(UV)/Mg~{\sc II} by more than a factor of two.
(\citealt{Goad2015} considered the effect of microturbulence on H$\beta$
emission and found that it increased emission across the whole BLR, but
especially at the smaller radii.) \citet{Verner2003} and \citet{Verner2004}
suggest that the most reasonable value for the microturbulence is
5--10~km~s$^{-1}$, while \citet{Bruhweiler2008} favour
20~km~s$^{-1}$. \citet{Ruff2012} suggest 100~km~s$^{-1}$ as that would
produce smooth line profiles. \citet{Sigut2003} and \citet{Baldwin2004} also
recognised that abundance is important but unlikely to be dominant.

Early in the history of the Fe~{\sc II} problem, \citet*{Wills1985} and
\citet{Collin-Souffrin1988} concluded that either there was an unusually high
abundance of iron or an important mechanism was being overlooked.
Microturbulence is one such mechanism since it increases the spread in
wavelength of Fe~{\sc II} absorption and thus increases radiative
pumping. Earlier than the discussions of microturbulence however,
\citet{Penston1987} had proposed that Ly$\alpha$ fluorescence might be the
overlooked mechanism. This possibility received observational and theoretical
support from \citet{Graham1996} and \citet{Sigut1998}
respectively. Ly$\alpha$ fluorescence of Fe~{\sc II} is discussed in detail
by \citet{Johansson1984} for cool stars and by \citet{Hartman2000} for the
symbiotic star RR~Tel.  These papers, and especially \citet{Hartman2000},
also discuss fluorescence arising from other emission lines, such as C~{\sc
  IV} $\lambda$1548.
\citet{Bruhweiler2008} discuss the significance of the peculiar atomic
structure of the Fe$^+$ ion. The 63 lowest energy levels, up to 4.77~eV, are
all of even parity, with no permitted transitions between them. These energy
levels will be well populated, with the electrons consequently available for
pumping to higher levels by the 10.2~eV Ly$\alpha$ line (and also by the
continuum).

\citet{Johansson1984} and \citet{Hartman2000} discuss Ly$\alpha$ fluorescence
in connection with stars, but the mechanism is, of course, equally relevant
to quasars. However, the width of Ly$\alpha$ is much larger in quasars than
in the stars. The most important excitation channels are within $\pm
3$~\AA\ of Ly$\alpha$, as discussed by \citet{Johansson1984}. For the broad
lines of quasars, \citet{Sigut1998} consider the excitation channels within
$\pm 50$~\AA, finding an increase of $\sim$~15 per cent in the ratio Fe~{\sc
  II}(UV)/H$\beta$ compared with that for $\pm 3$~\AA\ (with zero
microturbulence in both cases). We discuss briefly below
(section~\ref{subsecBLR}) that a particular small emitting region or cloudlet
will see the full profile of the Ly$\alpha$ arising from the whole ensemble
of cloudlets in the broad line region. For quasars, the central concentration
of the Ly$\alpha$ flux thus seems likely to be important, to emphasise the
important $\pm 3$~\AA. We might expect that, for quasars, the ratio of FWHM
to equivalent width of the Ly$\alpha$ would be a useful central-concentration
parameter for associating with the Fe~{\sc II} emission (with low FWHM/EW
implying a high central-concentration).

In this context of Ly$\alpha$ fluorescence, an interesting case is apparent
in \citet{Wills1980}. In 0957$+$561A and 0957$+$561B the UV Fe~{\sc II}
equivalent widths are different by a factor of nearly two, and the
differences are visually obvious in the spectra, although these objects are
gravitationally-lensed images of the same quasar. This observation suggests
that the Fe~{\sc II} emission can vary substantially on timescales comparable
to the $\sim$~1-year time delay between the two images \citep[more precisely,
  417 days][]{Kundic1997, Shalyapin2008}. The equivalent width of the nearby
Mg~{\sc II}~$\lambda$2798 emission and also of C~{\sc III]}~$\lambda$1909
  appeared unchanged. Iron abundance seems very unlikely to be the cause of
  the difference.  Ly$\alpha$ fluorescence is a possible cause of the
  difference, given that the Ly$\alpha$ flux in Q0957$+$561A,B has been
  observed to be variable on timescales (observed frame) of weeks
  \citep{Dolan2000}.

The continuum is substantially bluer in the image (A) in which the Fe~{\sc
  II} is weaker.  \citet{WillsWills1980} initially attributed this difference
in continuum shape to differential reddening along the different light-paths,
but, in a note added in proof, subsequently attributed it to the proximity of
lensing galaxy G1, at $\sim 1''$ from 0957$+$561B, in agreement with
\citet{Young1980} and then \citet{Young1981}. (Note that colour variability
has, however, been detected since by \citealt*{Shalyapin2012}.) The lensing
of 0957$+$561 ($z = 1.408$) arises from a cluster of galaxies ($z = 0.355$),
and G1, the brightest cluster member. G1 is about four times fainter in the
$R$ passband than 0957$+$561B (\citealt*{Walsh1979};
\citealt{Young1980}). The spectra of \citet{Wills1980} and
\citet{WillsWills1980}, which have much higher resolution and used smaller
apertures than the spectra of \citet{Young1980} and \citet{Young1981}, show
no indication of the $4000$\AA\ break from G1 affecting the spectrum of
0957$+$561B at $\sim 2250$\AA\ (rest frame), in the vicinity of the
low-wavelength end of the UV Fe~{\sc II} feature. The ratio $B/A$ of spectra
in Fig.~1 (single epoch) and Fig.~3 (combined epochs) of
\citet{WillsWills1980} appear consistent with the apparent differences in the
Fe~{\sc II} emission {\it not\/} being an artefact arising from G1.

\citet{Guerras2013} have investigated UV Fe~{\sc II} and Fe~{\sc III}
emission in the spectra of 14 image-pairs for 13 gravitationally-lensed
quasars. They find differences for four image-pairs, one of which is
Q0957$+$561A,B (apparent for Fe~{\sc III} only). They attribute these
differences to gravitational microlensing by stars in the lensing galaxies,
but caution that their result depends strongly on one image pair (SDSS
J1353$+$1138A,B). It is not clear why the explanation of the differences has
to be microlensing. From statistical considerations, they estimate that the
UV Fe~{\sc II} and Fe~{\sc III} emission arises in a region of size $\sim 4$
light-days and suggest that it is located within the accretion disk where the
continuum originates (size $\sim$~5--8 light-days). Possible implications of
this location for the existence of these ions and for the widths of emission
lines are not discussed.

\subsection{UV Fe~{\sc II} emission and the broad line region} \label{subsecBLR}

The UV Fe~{\sc II} emission is usually thought to arise in the broad line
region (BLR), but is sometimes attributed instead to an intermediate line
region (ILR), between the outer BLR and the inner torus. \citet{Graham1996}
and \citet{Zhang2011}, for example, favour the ILR interpretation.

Photoionisation equilibrium implies the temperature of the BLR gas is $\sim
10^4$~K. For such a temperature the thermal line-widths are
$\sim$~10~km~s$^{-1}$, which is very much smaller than the observed
line-widths of $\sim$~1000--20000~km~s$^{-1}$. Such a disparity might mean
that the BLR contains many ($\sim 10^8$) small cloudlets to produce an
overall smooth profile \citep{Dietrich1999}, although microturbulence will
also cause broadening \citep{Bottorff2000}, as will Rayleigh and Thomson
scattering \citep{Gaskell2013}. In reality, the distribution of the gas is
likely to be fractal \citep{Bottorff2001}.

The view of the structure of the BLR has changed somewhat over the years, in
substantial part because of the results of reverberation mapping of Seyfert
galaxies \citep[e.g.][]{Peterson2006}. In the old view there is a roughly
spherical distribution of the cloudlets, each with stratification of the
ionisation. Although much is uncertain, in the modern view there is a more
flattened distribution of cloudlets \citep{Gaskell2009, Kollatschny2011,
  Kollatschny2013}, with a general stratification of the ionisation and
density, such that high-ionisation and high density correspond to smaller
distances from the central black hole (BH) and low-ionisation and low density
correspond to larger distances \citep{Gaskell2009}. The flattening is more
pronounced for low-ionisation lines \citep{Kollatschny2013}. The BLR could be
a thick disk of cloudlets or could be bowl-shaped \citep*[][]{Goad2012,
  Pancoast2012, Pancoast2014}. In the RPC model \citep*[for radiation
  pressure confinement,][]{Baskin2014} the cloudlets are replaced by a BLR
that is a single stratified slab. High-ionisation lines tend to be broader
than low-ionisation lines in the same quasar. The low-ionisation optical
Fe~{\sc II} and Mg~{\sc II} emission are thought to arise in the same outer
regions \citep{Gaskell2009, Korista2004, Cackett2015}. There is some evidence
from reverberation mapping of Seyfert galaxies that the UV Fe~{\sc II} arises
at smaller radii in the BLR than the optical Fe~{\sc II} and is
correspondingly broader \citep[e.g.][]{Vestergaard2005, Barth2013}. The
flattened distribution rotates about the central BH, but there is also
(macro)turbulent motion of $\sim$~1000~km~s$^{-1}$ or more
\citep{Kollatschny2013}. Turbulence (macroturbulence) refers to bulk motion
measured orthogonal to the plane of rotation, but will be present in the
plane of rotation too \citep{Gaskell2009, Kollatschny2013}. Microturbulence
refers to (non-thermal) motion within cloudlets or between adjacent cloudlets
\citep{Bottorff2000}. The rotation velocity exceeds the turbulence velocity
by factors of a few times. Failure to account for the turbulence will lead to
the mass of the BH being overestimated \citep{Kollatschny2013}. In the case
of a BLR viewed face-on (i.e.\ viewing along the rotation axis, sometimes
also expressed as ``pole-on''), one is seeing mainly the turbulent motion in
the emission lines. The FWHM that we measure for the broad emission lines is
thus a function of viewing angle \citep{Wills1986}.

The amount of gas in the BLR exceeds by a large factor
($\sim$~$10^3$--$10^4$) that needed to account for the line emission. The
notion of ``locally optimally-emitting cloud'' (LOC), \citep{Baldwin1995} is
that the regions that are emitting strongly are those where the conditions to
do so are optimal. A change in the continuum luminosity can then result in an
apparent change of the scale-size of the BLR, simply because another region
is then the optimal emitter. From simple modelling of
  photoionisation, the scale-size should increase approximately as the
square-root of the ionising luminosity \citep{Peterson2006}. A particular
cloudlet or small region will receive the Doppler-shifted Ly$\alpha$ profile
of the whole BLR ensemble (i.e.\ broadened by the macroturbulent and
rotational velocities) but sliced by narrow absorption lines from intervening
cloudlets (Gaskell, private communication). (This fact is relevant to
Ly$\alpha$ fluorescence, as mentioned in section~\ref{subsecFeII}.) The
cloudlet itself will be emitting with a line-width comparable to the sound
velocity or microturbulence velocity.

The investigation of metallicity in the nuclear regions of quasars proceeds
by analysis of the broad emission lines and intrinsic (i.e.\ associated with
the quasar, not intervening) narrow absorption lines
\citep[e.g.][]{Hamann1999, Hamann2004}. The growth of the central
supermassive BH is believed to arise from accretion following large-scale
events in the galaxies --- mergers and interactions --- and from quieter,
secular processes such as flows along bars.
Infall of (possibly pristine) gas might also be involved. The gas that
approaches the nucleus either forms stars or it settles into the dusty torus,
and from there spirals into the accretion disk and the BH. The distinction
between the torus and the accretion disk is that the temperature of the
accretion disk exceeds the sublimation temperature of the dust. The BLR is
actually the turbulent gas above the accretion disk, and which rotates with
it. (Boundaries between accretion disk, torus and BLR are related
  to the cooling process that is dominant: continuum-cooling for the
  accretion disk; thermal emission from dust for the torus; line-cooling for
  the BLR.)
Some small fraction of the incoming gas is expelled in a wind that removes
angular momentum and allows the accretion. Dynamical models of
  the BLR suggest inflows, outflows, and orbital motion \citep[e.g.][]{
    Pancoast2012, Pancoast2014, Grier2015}. The metallicity of the BLR
should be that of the torus.
\citet{Czerny2011} discuss, and give a good illustration in their Figure 1,
the accretion disk extending from the BH to the outer edge of the torus,
where the torus dust sublimates, with the clouds (low-ionisation clouds at
least) of the BLR ``boiling'' (inflowing and outflowing) from its outer
parts. In their interpretation the atmosphere of the outer parts of the disk
is cool enough for dust to be present. The outflow of the boiling is driven
locally by the accretion disk --- a dust-driven wind ---, but the driving
force is cut-off at large heights by the central radiation sublimating the
dust, leading to infall. Outflow and inflow would lead to turbulence.
Note, however, that, for NGC~4151, \citet{Schnuelle2013} find evidence that
the inner torus is actually located beyond the sublimation radius and that it
does not have a sharp boundary \citep[see also][]{Kishimoto2013}.

\citet{Hamann2004} consider the bright phase of a quasar to correspond to a
final stage of accretion during which the BH roughly doubles its mass. This
phase lasts for $\sim 6 \times 10^7$~yr for accretion at $\sim$~50 per cent
of the Eddington rate. The metallicities of the BLR and the torus can be
expected to be typical of the central parts of galaxies at the redshifts
concerned. The more massive galaxies can be expected to have higher central
metallicities because of repeated episodes of star formation, and because the
deeper potential well increases the retention of gas for recycling. The
observations indicate nuclear metallicities of solar to several times
solar. It is also possible that mergers and interactions could drive
relatively unprocessed gas into the nuclear regions, which would reduce the
metallicity. Radiation pressure could, in principle, concentrate metals
relative to hydrogen in some regions (e.g.\ in BAL outflows or in the outer
parts of accretion disks), but observationally the effect is not currently
known to occur \citep{Baskin2012a, Baskin2012b}. Quasar activity is believed
to be preceded by star-formation activity. One might then expect to see a
dependence on redshift of the metallicity of quasars, but none has so far
been detected for $z \la 6.5$. As mentioned in section~\ref{subsecFeII},
enrichment of the BLR appears to be completed well before (0.3--1~Gyr) the
visible quasar activity \citep[e.g.][]{Dietrich2003, Simon2010,
  DeRosa2011}. There is no indication of further enrichment from continuing
star formation.

\subsection{Index of the UV Fe~{\sc II} strength: ${\bf W2400}$}
\label{subsecIndexW2400}

As the index of UV Fe~{\sc II} emission, we again (as in Clo13b) use the
rest-frame equivalent width $W2400$ \citep{Weymann1991}, which is defined
between two continuum-windows, 2240--2255 and 2665--2695~\AA, and integrated
between 2255--2650~\AA. A very similar approach is taken by
\citet{Sameshima2009}. It is helpful to define some criteria for ``strong'',
``ultrastrong'', and ``weak'' emitters. In Table~2 of \citet{Weymann1991} the
median $W2400$ is $\sim$~30~\AA\ for all quasars (non-BAL and BAL). We then
define (as in Clo13b) ``strong emitters'' as those with $30 \le W2400 <
45$~\AA, ``ultrastrong'' as $W2400 \ge 45$~\AA, and ``weak'' as $W2400 <
30$~\AA. Also, in this paper, we sometimes make use of sub-divisions of the
weak category into upper, middle, and lower thirds: weak\_upper with $20 \le
W2400 < 30$~\AA; weak\_middle with $10 \le W2400 < 20$~\AA; weak\_lower with
$W2400 < 10$~\AA.

\subsubsection{$W2400$ and the 2175~\AA\ dust feature}
\label{subsubsecDust2175}

Note that the range of wavelengths corresponding to $W2400$ is overlapped by
the broad 2175~\AA\ dust feature. This 2175~\AA\ feature is characteristic of
extinction in the Milky Way (MW), is present but weaker in the LMC, and is
apparently not present in the SMC \citep{Mathis1990}. Its origin is usually
ascribed to grains of carbon and polycyclic aromatic hydrocarbons
\citep[PAHs; e.g.][]{Hecht1986, Draine2003}. The strength of the
2175~\AA\ feature in the MW might be somewhat anomalous. It appears to be
present only rarely in other galaxies, although investigation is
difficult. There is little variation of the central wavelength of
2175~\AA\ \citep{Mathis1990}, but the FWHM can reach, on the long-wavelength
side, from $\sim$ 2370 to $\sim$ 2640~\AA\ \citep{Mathis1990}, and typically
$\sim 2440$~\AA\ \citep{Draine1989}.
Thus, the 2175~\AA\ feature, if present, could overlap from a little to most
of the $W2400$ range. If it affected the lower continuum-window but not the
upper then the setting of the continuum could be too low and the $W2400$
measurement too large.

In fact, there appears to be no compelling evidence that the
2175~\AA\ feature does affect the spectra of quasars significantly.
\citet*{Pitman2000} concluded, in a review, that quasars seem to have
SMC-type extinction and there were no real detections of the
2175~\AA\ feature in quasars. Subsequently, \citet{Hopkins2004} concluded
from a large sample of SDSS quasars that the reddening in the ``red tail'' of
the colour distribution is SMC-like. Note that for most quasars the carbon
grains and PAHs would probably be destroyed by the X-ray to ultraviolet
photons. \citet{Zhang2015} find a feature at $\sim 2250$~\AA\ (``EBBA'' ---
excess broad-band absorption) in the spectra of seven BAL quasars from
SDSS-III~/~BOSS~/~DR10 \citep{Paris2014} and tentatively raise the
possibility that it might be related to the 2175~\AA\ feature. BALs could
shield the dust from the photons that would destroy it. (Compare with cases
in which a claimed 2175~\AA\ feature is associated not with the quasars but
with the galaxies causing intervening absorption in the quasar spectra ---
e.g. \citealt{Noterdaeme2009, Jiang2010}.)  \citet{Zhang2015} say they have
18 further candidates (non-BAL) for the 2175~\AA\ feature from the DR10
spectra. A possible concern for our work here would be if the
2175~\AA\ feature was preferentially present in the lower-luminosity quasars:
the fact that Zhang et al.\ are finding so few candidates in the typically
fainter DR10 quasars suggests that it is not.

\subsection{Intrinsic continuum luminosity: ${\bf L3000}$}
\label{subsecIntContFlux}

For the measure of intrinsic luminosity, we use $L3000$, which is defined as
the intrinsic continuum flux in units of erg~s$^{-1}$ \citep[e.g.][]{Shen2011}:
$$L3000 = \lambda f_\lambda(\lambda) \cdot 4\pi d_L^2 {\rm ,}$$ where
rest-frame $\lambda \equiv 3000$~\AA\, and $d_L$ is the luminosity
distance. We measure $L3000$ as the median across $100$~\AA\ centred on
$3000$~\AA.

Measures for other wavelengths may be defined similarly. We make a little use
also of $L2200$.

\section{Measuring the spectra: ${\bf W2400}$, ${\bf W2400\MakeLowercase{g}}$,
${\bf W2798}$, ${\bf L3000}$, and other quantities} \label{secMeasuring}

We use software to measure the Fe~{\sc II} rest-frame equivalent width,
$W2400$, essentially as described in Clo13b, but we now include some
additional measurements: (i) a measure of the gradient, $W2400g$, of the
continuum local to the measurement of $W2400$, expressed as a colour; (ii)
the median signal-to-noise ratio of each spectrum, $sn\_med$; (iii) the
rest-frame equivalent width of the Mg~{\sc II} emission, $W2798$, calculated
in a similar way to $W2400$; and (iv) the FWHM of the Mg~{\sc II} emission,
$fwhm2798$. In Clo13b, for which we were using only SDSS DR7QSO spectra, we
used the SDSS s/n measures in the FITS headers.

$W2400g$ is a measure of the gradient of the line taken as the continuum in
the \citet{Weymann1991} definition of $W2400$. We represent it as a colour,
calculated as $W2400g = -2.5\log_{10}(l_1/l_2)$, where $l_1$ and $l_2$ are
the median flux densities from the first and second continuum windows
(2240--2255, 2665--2695~\AA). In this way, $W2400g$ is: zero colour for a
flat continuum line; negative or blue colour for a continuum line increasing
to bluer wavelengths; and positive or red colour for a continuum line
increasing to redder wavelengths.

An alternative measure of the local gradient of the continuum, or continuum
colour, could be $-2.5\log_{10}[(L2200/2200)/(L3000/3000)]$, where $L2200$ is
defined similarly to $L3000$.

The s/n, $sn\_med$, is calculated from blocks of 65 pixels running through
each (unsmoothed) spectrum. That is, a given block of 65 pixels is centred
one pixel beyond the preceding block. For a given block, the ``signal'' is
taken as the median flux density, and the ``noise'' is taken as the median
absolute deviation from the signal. $sn\_med$ is then taken as the median
signal / noise of all such blocks, excluding those affected by the edges of
the spectra. It should not be biased by the better s/n in the relatively
small regions of the spectra occupied by emission lines. It is a different
measure from those found in the headers of the SDSS spectra. The intention is
not that $sn\_med$ is of any particular astrophysical interest, but that it
gives a robust and general indication of the quality of the data that can be
applied consistently to spectra from different sources (here, SDSS and MMT /
Hectospec spectra). It allows a threshold to be set at which we can take the
quantities that are of astrophysical interest to be determined reliably.

In our subsequent processing, all spectra are first smoothed with a 5-pixel
median filter, for the purpose of setting the continuum levels more reliably,
given the quite narrow windows of the \citet{Weymann1991} definition of
$W2400$.

\begin{figure*}
\includegraphics[height=80mm]{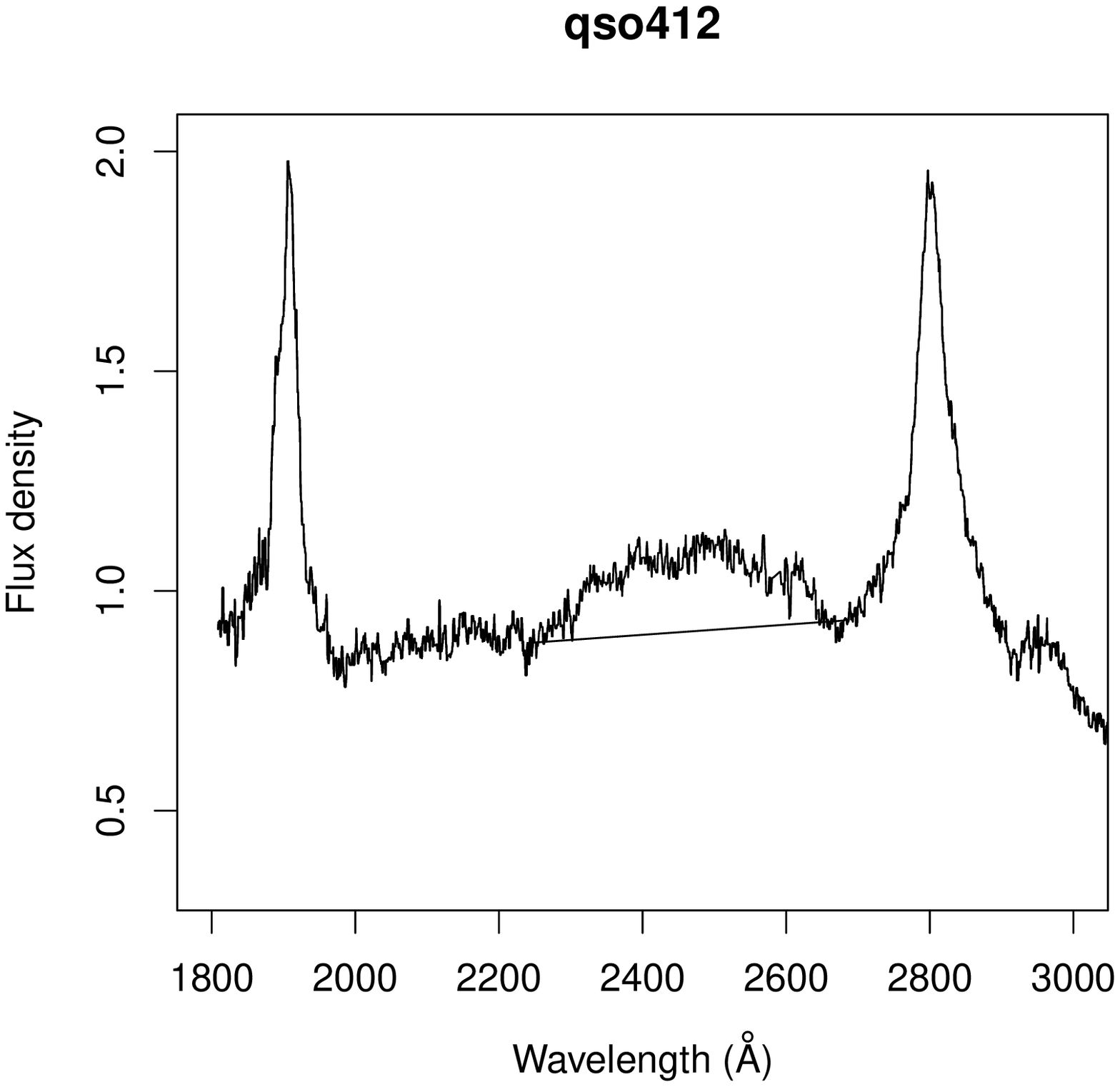}
\includegraphics[height=80mm]{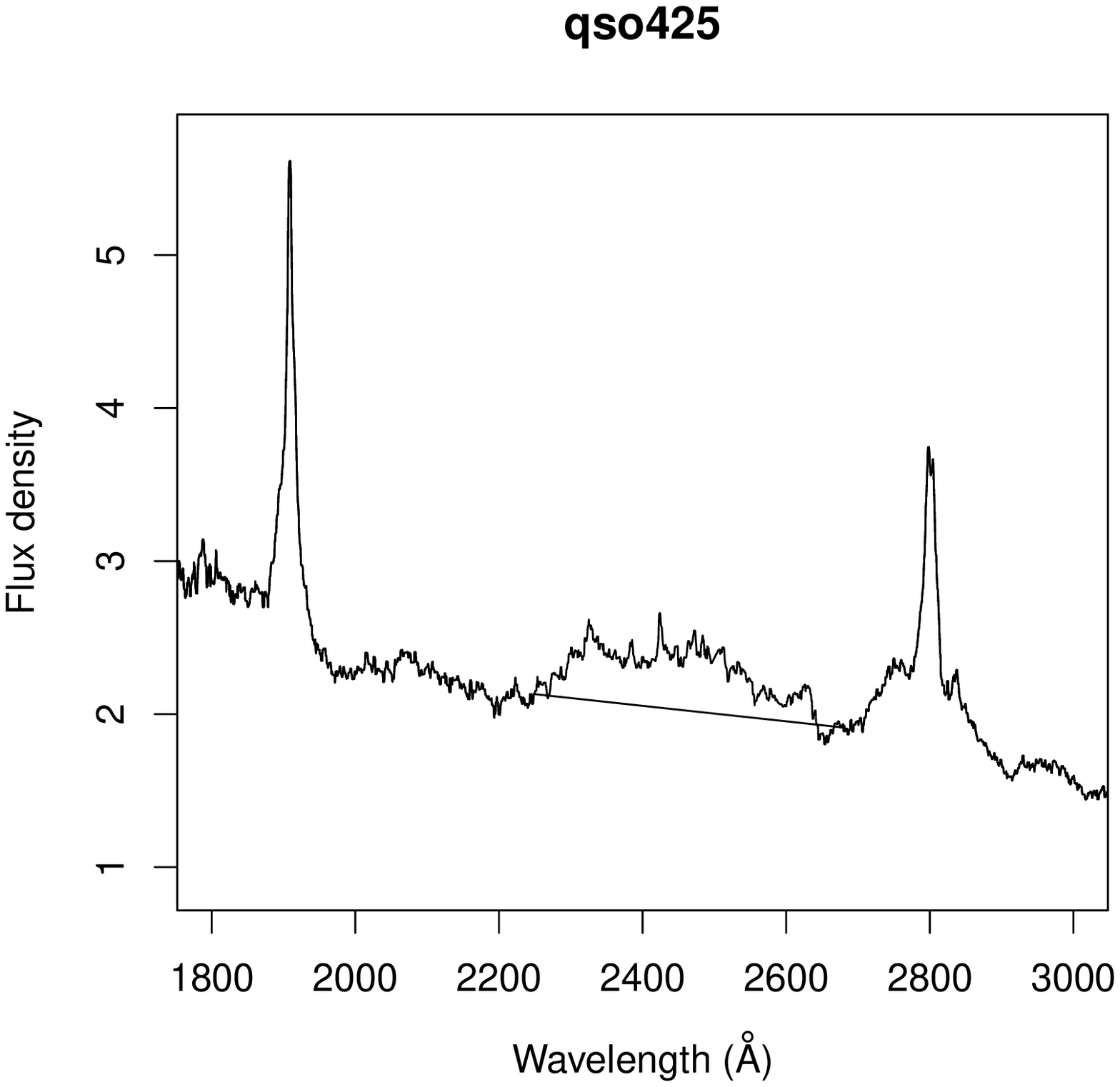}

\caption{
Rest-frame (optical) spectra of two ultrastrong UV Fe~{\sc II}-emitting
quasars, qso412 ($z = 1.156$, $W2400 = 55.5$~\AA) and qso425 ($z = 1.230$,
$W2400 = 52.5$~\AA), from our MMT / Hectospec spectroscopy
(Appendix~\ref{secAppendixMMTHectospecGALEX}). The straight-line section of
each spectrum illustrates the continuum between the two continuum
windows. The rest-frame equivalent width $W2400$ is calculated with respect
to this continuum \citep{Weymann1991}. The spectra have been smoothed with a
5-pixel median filter. The flux densities ($f_\lambda$) are in units of
$10^{-17}$~erg~s$^{-1}$~cm$^{-2}$~\AA$^{-1}$. The wavelength range has been
restricted to $\sim$~1800--3000~\AA\ for clarity. See
Table~\ref{mmt_hecto_quasars} for further details of these two quasars.
}
\label{w2400_example_spectra}
\end{figure*}

As mentioned in section~\ref{subsecIndexW2400}, the rest-frame equivalent
width $W2400$ \citep{Weymann1991} is defined between two continuum-windows,
2240--2255 and 2665--2695~\AA. For each continuum window, the median flux
density representing the continuum is attributed to the centre of the window.
The integration for the equivalent width is between 2255--2650~\AA. Note that
the Fe~{\sc II} flux will be low but not zero in these two continuum-windows:
one cannot expect to obtain the ``true continuum'' in this region of the
spectrum. The continuum between the two windows is taken to vary linearly.
With the median filtering, Clo13b estimated indicative errors in $W2400$ of
$\sim$~3.5--3.7~\AA\ for SDSS DR7QSO spectra with $i \le 19.1$ and $1.0 \le z
\le 1.8$. Fig.~\ref{w2400_example_spectra} illustrates the continuum between
these two continuum windows for two ultrastrong emitters from our own
data sets, in this case MMT / Hectospec spectroscopy
(Appendix~\ref{secAppendixMMTHectospecGALEX}).

In this Fig.~\ref{w2400_example_spectra}, clearly qso412 has a redder
continuum local to the measurement of $W2400$ than qso425. If we artificially
change the curvature of the spectrum of qso412, by a $1/\lambda$ function, to
resemble that of qso425 we find that $W2400$ is reduced from 55.5~\AA\ to
52.1~\AA. The difference is comparable to the indicative measurement
errors. Full details of the $1/\lambda$ function are given in
section~\ref{secFurtherNotes}, where we discuss further the effects of
curvature of the spectra.

One small change compared with Clo13b is that we now assume there will be a
residual [O~{\sc I}] $\lambda$5577~\AA\ sky feature and interpolate across
it. Previously, we simply allowed the median filtering to reduce any residual
sky feature. In most cases its effect then, if it coincided with the Fe~{\sc
  II} emission, was smaller than the indicative errors. Also, in the case of
our MMT / Hectospec spectra only (i.e.\ not the SFQS spectra and not the SDSS
spectra --- Appendix~\ref{secAppendixMMTHectospecGALEX}), the atmospheric
A-band at $\sim$~7600~\AA\ is prominent and we have interpolated across it
too.

The use of $W2400$ allows consistent measurements across different data
sets. Possibly the fitting of a template to the Fe~{\sc II} could restore
some information that might presently be lost to uncertainties in measuring
$W2400$. However, \citet{Verner2004} say that a programme of both
observations and modelling would be needed to establish the validity of a
single template.
\footnote{
\citet{Vestergaard2001} provide a single template, derived from observations
of the low-redshift Seyfert~1 galaxy I~Zw~1. Although Vestergaard \& Wilkes
successfully applied their template to four other quasars, they were
themselves cautious about its wider application. In particular, Vestergaard
\& Wilkes note the two assumptions about the use of the template: (i) that
the iron spectrum of I~Zw~1 is representative; and (ii) that the iron spectra
of quasars in general can be fitted simply by scaling and broadening the
template. Furthermore, Vestergaard \& Wilkes advise that the fitting process
should divide the template into segments to be fitted independently, because
the scaling factors relative to the continuum are different for different
segments. (Their recommended fitting process is also iterative and involves
visual checking and manual intervention.)

If the template-fitting is going to fail then we might expect the failure to
be most apparent for the ultrastrong / strong emitters.  \citet{Shen2011}
have used the \citet{Vestergaard2001} template on the same DR7QSO spectra
(with an automated fitting process) and so we have an opportunity to make
comparisons between the two approaches. The Weymann et al.\ $W2400$ measure
is integrated across 2255--2650~\AA. The Shen et al.\ template measure of the
UV~EW~Fe ($EW\_FE\_MGII$) is measured on both sides of the Mg{\sc II}, across
2200--3090~\AA\ (probably meaning 2200--2700~\AA\ and 2900--3090\AA\ but this
is unclear from their paper). Nevertheless, the two measures should
correlate. We restrict comparisons to those quasars in the DR7QSO catalogue
that have $ 1.0 \le z \le 1.8$, $i \le 19.1$ and that satisfy our s/n
criterion, $sn\_med \ge 10.0$. We find that the two measures do correlate,
especially for our weak category of emission, but that, as the $W2400$
measure moves into our strong and ultrastrong categories, the Shen et
al.\ measure deviates to increasingly relatively large values. Eventually,
the Shen et al.\ measures seem to become implausibly large. We conclude that
the $W2400$ measure is to be preferred. }

The rest-frame equivalent width $W2798$ for Mg~{\sc II} emission is
calculated, in a similar way to $W2400$, for continuum windows 2670--2700 and
2900--2930~\AA, and integrated between 2700--2900~\AA. Note that the Mg~{\sc
  II} emission can be affected by Fe~{\sc II} emission, especially on the
blueward side.

We also calculate a simple measure of the FWHM of the Mg~{\sc II} emission by
subtracting the continuum, applying a 19-pixel median filter (in addition to
the existing 5-pixel median filtering) to produce a smoother line-profile,
and calculating the resulting FWHM, $fwhm2798$. Median filtering preserves
edges and does not itself introduce broadening. The filtering of the maximum,
reducing it, will lead to the FWHM being a little larger than it should be,
but the effect should be generally consistent and equivalent to not half of
the true maximum but a slightly smaller fraction of it. We have not made any
correction for a component of Mg~{\sc II} that might arise from the
narrow-line region (FWHM $< 1200$~km~s$^{-1}$): both \citet{Shen2008} and
\citet{Shen2011} indicate that there is no clear need to do so. Typically,
the measurement of $fwhm2798$ will be a measurement of the core of the
Mg~{\sc II} emission line, and so not substantially affected by any Fe~{\sc
  II} emission that might be present in the wings. We suspect that some of
the FWHM measurements from \citet{Shen2011} have been affected by Fe~{\sc II}
emissions in the wings\footnote{We note that our measure of FWHM, $fwhm2798$,
  correlates well with the $FWHM\_MGII$ measure (whole profile ---
  i.e.\ broad $+$ narrow) from \citet{Shen2011}, and $fwhm2798 \approx
  FWHM\_MGII$. It also correlates well with their preferred
  $FWHM\_BROAD\_MGII$ measure (broad profile --- i.e. narrow-subtracted) but
  there is an asymmetry in the scatter about $fwhm2798 = FWHM\_BROAD\_MGII$,
  such that $FWHM\_BROAD\_MGII$ is relatively larger for middle-of-the-range
  values. We note also that a plot of their $FWHM\_MGII$ (whole profile)
  against their $FWHM\_BROAD\_MGII$ (broad profile) has a complicated
  structure. We suspect, therefore, that subtraction of a narrow component
  (regardless of whether it arises from the NLR) can lead to residual (after
  their subtraction of an iron template) neighbouring Fe~{\sc II} being
  wrongly attributed to broad Mg~{\sc II}.}.

We measure $L3000$ from the median flux density of the smoothed spectra in
the wavelength range 2950--3050~\AA. We have not corrected the spectra for
Galactic extinction (i.e.\ our Galaxy).

Measurement by software has great advantages compared with interactive
measurements. The \citet{Weymann1991} method can be applied objectively to
each spectrum, and a large number of spectra can be processed. As discussed
in Clo13b, approximately 1 per cent of the measurements of $W2400$ by the
software are negative. Most commonly, negative $W2400$ occurs with flux
densities that are increasing rapidly to shorter wavelengths, leading to
concave spectra. Rigorous application of the method to concave spectra will
lead to negative $W2400$. Occasionally, negative $W2400$ occurs because of
absorption, artefacts, and, for $W2400 \approx 0$, noise fluctuations.
Interactive measurements, in contrast, struggle to apply the method
objectively and consistently, and they cannot easily process many
spectra. They can, however, easily interpolate for the small fraction of
spectra that are affected by absorption and artefacts.

We use the following wavelength ranges in the processing of spectra: SDSS
spectra, 3800--9200~\AA; SFQS MMT / Hectospec spectra, 3800--8500~\AA;
our MMT / Hectospec spectra, 3900--8200~\AA.

\section{UV F\lowercase{e}~{\sc II} in fainter quasars}
\label{secFainter}

In this section we investigate the UV Fe~{\sc II} properties of fainter
quasars relative to brighter quasars, for the redshift range $1.0 \le z \le
1.8$. We begin with the SDSS DR7QSO catalogue and spectra
\citep{Schneider2010}. The limit $i \le 19.1$ was specified for the
low-redshift strand of quasar selection \citep{Richards2006,
  Vanden-Berk2005}. The DR7QSO catalogue contains 105783 quasars, of which
43604 have $1.0 \le z \le 1.8$; of these 43604, 27991 have $i \le 19.1$. For
these quasars, we can distinguish between quasars that are members of LQGs
and those that are not \citep[as in Clo13b,][]{Clowes2013a, Clowes2012} using
our main DR7QSO catalogue of LQG candidates (Clowes, in preparation).

We extend the discussion to fainter quasars using the SFQS catalogue and
spectra \citep{Jiang2006}. The SFQS catalogue contains 414 quasars, reaches
$g \sim 22.5$, and covers $\sim$~3.9~deg$^2$ of the SDSS stripe~82. There are
five sub-fields, in four non-contiguous patches. Spectroscopy is from MMT /
Hectospec for 366 of the quasars and from the SDSS for the remaining 48. Of
the 414, 178 are in the redshift range $1.0 \le z \le 1.8$ of interest here;
of these 178, 90 have $g \le 21.0$.

We have used the DR7QSO data to assess whether we can find LQGs that could
affect the SFQS data for $1.0 \le z \le 1.8$, given the limitations of the
narrowness of stripe~82 and a brighter magnitude limit. From our main DR7QSO
catalogue of LQG candidates ($i \le 19.1$, $1.0 \le z \le 1.8$, linkage scale
of 100~Mpc, minimum membership of 10 quasars) there is a small overlap of the
SFQS with a LQG candidate of 12 members at $\bar{z} = 1.087$ ($z$-range:
1.0444--1.1356), but there are no quasars in common. Nevertheless, a small
redshift spike (6 with $z$: 1.00--1.04) in this patch of the SFQS data does
suggest that the LQG is perceptible in these mostly fainter quasars. Its
influence on our statistical analyses and conclusions should be
negligible. Specifically for this discussion of the SFQS, we have also
produced a DR7QSO catalogue of LQG candidates for stripe~82 only (no limit on
$i$, $1.0 \le z \le 1.8$, linkage scale of 54~Mpc, minimum membership of
eight quasars), to allow for its greater depth, but we find no indications of
further LQGs that could affect the SFQS data.

Another possibility for a faint quasar survey with a large telescope is 2SLAQ
\citep{Croom2009}, but we found it to be less suitable than SFQS because the
data are not flux-calibrated, are given as counts, are not corrected for the
response function, and have generally poorer s/n. While the second of these
problems is easily addressed the other three are not.

We first use a subset of these DR7QSO quasars as a high-s/n reference sample,
{\sc dr7qso15}, with the ``15'' indicating the limit to $sn\_med$, with which
to investigate the consequences of decreasing s/n. The full definition of
{\sc dr7qso15} is: $1.0 \le z \le 1.8$, $i \le 19.1$, and $sn\_med \ge
15.0$. It has 15131 quasars --- the DR7QSO catalogue is so large that even
such a stringent limit on $sn\_med$ gives a large sample. From the s/n values
in the SDSS headers, $sn\_med \ge 15.0$ is roughly equivalent to $sn\_g \ga
10.0$, $sn\_r \ga 12.3$, $sn\_i \ga 10.4$, and $sn\_worst \ga 9.6$, where
$sn\_worst$ is, for any spectrum, the lowest of the three SDSS s/n measures.

As s/n decreases, generally as magnitude increases, the dispersion of the
$W2400$ measurements becomes larger. An important question when, comparing
distributions with the Mann-Whitney test \citep[e.g.][]{DeGroot2012}, is
whether decreasing s/n leads not only to increased dispersion but also to a
systematic shift of $W2400$. By adding gaussian random noise to artificially
degrade the s/n for the spectra of {\sc dr7qso15} we find that no systematic
shift is introduced, provided that the $W2400$ measurements are allowed to
become negative. (It is possible that interactive measurements would
subconsciously constrain the $W2400$ measurements to be always positive.)
This result, that degrading the s/n increases the dispersion but does {\it
  not\/} introduce a systematic shift of the $W2400$ distribution, is
important for what follows.

Subsequently, from the DR7QSO catalogue, we shall use the subset {\sc
  dr7qso10} for which the definition is: $1.0 \le z \le 1.8$, $i \le 19.1$,
and $sn\_med \ge 10.0$. It has 25742 quasars. The s/n limit ${\rm
  sn\_med} \ge 10.0$ is roughly equivalent to SDSS $sn\_g \ga 6.7$,
$sn\_r \ga 8.8$, $sn\_i \ga 7.8$, and $sn\_worst \ga 6.6$.

Similarly, from the SFQS catalogue, we shall use the subset {\sc sfqs10}
for which the definition is: $1.0 \le z \le 1.8$ and $sn\_med \ge
10.0$.  It has 83 quasars. Note that the SFQS has no specified limit on the
magnitude that can be taken as the counterpart of $i \le 19.1$ for the DR7QSO
catalogue.

The definitions of these, and subsequent samples, are summarised in
Table~\ref{sample_table_1}.

\begin {table*}
\flushleft
\caption {A summary of the definitions of the quasar samples. The columns are
as follows.  (1)~Name of the sample.  (2)~Source from which the sample is
derived.  (3)~Range in the $sn\_med$ s/n of the spectra ($sn\_med$ is
described in the text).  (4)~Range in redshift $z$.  (5)~Range in magnitude.
  (6)~Size $n$ of the sample.  (7)~Notes.
}
\small \renewcommand \arraystretch {0.8}
\newdimen\padwidth
\setbox0=\hbox{\rm0}
\padwidth=0.3\wd0
\catcode`|=\active
\def|{\kern\padwidth}
\newdimen\digitwidth
\setbox0=\hbox{\rm0}
\digitwidth=0.7\wd0
\catcode`!=\active
\def!{\kern\digitwidth}
\begin {tabular} {lllllll}
\\
(1)                              & (2)              & (3)                & (4)                 & (5)                   & (6)   & (7)                          \\
Sample                           & Source           & sn\_med range      & z range             & mag.\ range           & n     & Notes                        \\
                                 &                  &                    &                     &                       &       &                              \\
\\
{\sc dr7qso15}                   & DR7QSO           & $sn\_med \ge 15.0$ & $1.0 \le z \le 1.8$ & $i \le 19.1$          & 15131 &                              \\
\\
{\sc dr7qso10}                   & DR7QSO           & $sn\_med \ge 10.0$ & $1.0 \le z \le 1.8$ & $i \le 19.1$          & 25742 &                              \\
\\
{\sc sfqs10}                     & SFQS             & $sn\_med \ge 10.0$ & $1.0 \le z \le 1.8$ &                       &    83 & No specified magnitude limit \\
\\
\end {tabular}
\\
\label{sample_table_1}
\end {table*}

\subsection{Fainter quasars and ${\bf W2400}$} \label{subsecFainterW2400}

Recall that Clo13b showed that the distribution of $W2400$ for quasars that
are members of LQGs is shifted to larger values compared with that for
non-members, matched in magnitude and redshift. That is, the shift in the
$W2400$ distribution was a {\it differential\/} effect, for LQG members with
respect to non-members. There was a tentative indication that the size of
this differential shift increased with the $i$ magnitude of the
quasars. Also, the differential shift appeared to be strongly concentrated in
the redshift range $1.1 \le \bar{z}_{LQG} \le 1.5$. The matched control
sample used there was intended to negate any universal magnitude and redshift
dependences (i.e.\ not depending on environment) that might arise from, for
example, a hypothetical universal increase of $W2400$ with decreasing
intrinsic continuum luminosity. Here, we elaborate on the luminosity
dependences, in terms of the intrinsic continuum flux $L3000$, for both
quasars in general (universal) and for quasars that are members of LQGs
(differential).

\begin{figure*}
\includegraphics[height=80mm]{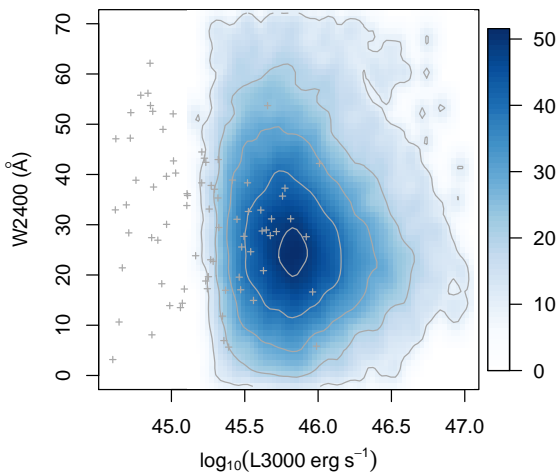}
\caption{A plot of $W2400$ against $\log_{10}L3000$ for the sample {\sc
dr7qso10}. $L3000$ is the intrinsic continuum flux, in units of
erg~s$^{-1}$, at 3000~\AA\ in the rest frame (see the text for its
definition). The shading indicates the kernel-smoothed densities of points
in a $64 \times 64$ grid, because the number of points, 25700, is too high
for an ordinary scatterplot to be a useful illustration. Linear contours
for the densities are also shown. The plot has been restricted to $44.6 \le
log_{10}L3000 \le 47.0$ and $0 \le W2400 \le 70$~\AA\ for clarity. Note
that the highest Fe~{\sc II} emitters tend to favour lower values of
$\log_{10}L3000$. The plot also shows, as points (crosses), $W2400$ against
$\log_{10}L3000$, for the sample {\sc sfqs10} (80 quasars). The same trend
is apparent for the {\sc sfqs10} quasars.
}
\label{LOG_L_3000_w2400_F1_c}
\end{figure*}

For 25700 quasars (from 25742) of the sample {\sc dr7qso10}, and for 80
quasars (from 83) of the sample {\sc sfqs10}, we can measure all of $W2400$,
$fwhm2798$, and $L3000$. In Fig.~\ref{LOG_L_3000_w2400_F1_c} we plot, for
{\sc dr7qso10}, $W2400$ against $\log_{10}L3000$. Because 25700 points is too
high for an ordinary scatterplot to be a useful illustration the plot shows
instead the kernel-smoothed densities of points in a $64 \times 64$
grid. Linear contours for the densities are also shown. For {\sc dr7qso10},
the plot shows that the highest Fe~{\sc II} emitters tend to favour lower
values of $\log_{10}L3000$. Note, however, that it is not an exclusive
relation: high Fe~{\sc II} emission does not guarantee low $\log_{10}L3000$,
and low $\log_{10}L3000$ does not guarantee high Fe~{\sc II} emission. In
this figure we also plot, as points, $W2400$ against $\log_{10}L3000$, for
sample {\sc sfqs10} (80 quasars), which is generally a lower-luminosity
sample. The same trend is apparent for the {\sc sfqs10} quasars.

\begin{figure*}
\includegraphics[height=80mm]{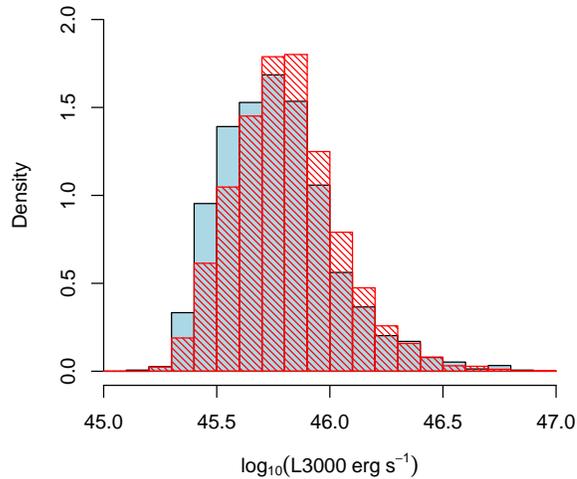}
\caption{The distributions of $\log_{10}L3000$ for the 1531 ultrastrong
$W2400$ emitters of the {\sc dr7qso10} sample (solid histogram, blue
online) and for the 7086 strong $W2400$ emitters (hatched histogram, red
online). $L3000$ is the intrinsic continuum flux, in units of erg~s$^{-1}$,
at 3000~\AA\ in the rest frame (see the text for its definition). Both are
density histograms. The bin size is 0.1 in the logarithm. The histograms
have been restricted to $45.0 \le log_{10}L3000 \le 47.0$~\AA\ for
clarity. The distribution for the ultrastrong emitters is clearly shifted
to lower values of $\log_{10}L3000$ relative to that for the strong
emitters.
}
\label{LOG_L_3000_F1}
\end{figure*}

This tendency of the highest Fe~{\sc II} emission favouring relatively low
$\log_{10}L3000$ is further illustrated in Fig.~\ref{LOG_L_3000_F1}, which
shows, for the {\sc dr7qso10} sample, the distributions of $\log_{10}L3000$
for the 1531 ultrastrong emitters (solid histogram, blue online) and for the
7086 strong emitters (hatched histogram, red online). The difference in the
distributions is very clear.

A one-sided Mann-Whitney test indicates that there is a shift to smaller
values of $\log_{10}L3000$ for the ultrastrong emitters (1531 quasars)
relative to the strong emitters (7086 quasars) at a level of significance
given by a p-value of $1.260 \times 10^{-12}$. The median shift is estimated
as $0.046 \pm 0.006$ in the logarithm, using the Hodges-Lehmann estimator
\citep{Hodges1963}. In contrast, there is no relative shift in the
distributions of $\log_{10}L3000$ for the two weaker intervals $20 \le W2400
< 30$ (weak\_upper, 9845 quasars) and $10 \le W2400 < 20$ (weak\_middle, 5871
quasars). They are indistinguishable (p-value of $0.504$).

These and other results in this section from our applications of the
one-sided Mann-Whitney test are summarised in Table~\ref{mw_summary_table}.
Note that we have avoided comparisons with the interval weak\_lower because
the measurements of $W2400$ there can become comparable to the indicative
errors.

Similarly, for the smaller but generally lower-luminosity sample {\sc
  sfqs10}, a one-sided Mann-Whitney test indicates that there is a shift to
smaller values of $\log_{10}L3000$ for the ultrastrong emitters (12 quasars)
relative to the strong emitters (30 quasars) at a level of significance given
by a p-value of $3.545 \times 10^{-4}$. The median shift is estimated as
$0.376 \pm 0.093$ in the logarithm. Similarly, again, there is no relative
shift in the distributions of $\log_{10}L3000$ for the two weaker intervals
$20 \le W2400 < 30$ (weak\_upper, 17 quasars) and $10 \le W2400 < 20$
(weak\_middle, 15 quasars). They are again indistinguishable (p-value of
$0.892$).

Note that this trend for $W2400$ to increase as $L3000$ (or $\log_{10}L3000$)
decreases is not strictly a correlation --- it does not exist across the
entire range of $W2400$ and $\log_{10}L3000$. From the large {\sc dr7qso10}
sample, the trend exists for $W2400 \ga 25$~\AA, but $W2400$ appears to be
essentially independent of $\log_{10}L3000$ for $W2400 \la 25$~\AA\ --- see
again the contours in Fig.~\ref{LOG_L_3000_w2400_F1_c}. The observed trend
should therefore not be confused with a Baldwin effect\footnote{We note that
  \citet*[][Fig.~1, for $L_{2500}$]{Green2001} and \citet[][Fig.~7, for
    $L1450$]{Dietrich2002} have previously reported, as Baldwin effects,
  associations of the equivalent width of UV Fe~{\sc II} with intrinsic
  luminosity. Both results are for wide ranges of redshift. \citet{Green2001}
  suspect that their effect is more strongly dependent on redshift than
  luminosity. \citet{Dietrich2002} cautiously qualify their effect with the
  phrase ``\ldots\ some indications \ldots''. \citet{Shen2011} appear not to
  discuss any possible association of UV Fe~{\sc II} and intrinsic
  luminosity.}.

If we consider the quasars from {\sc dr7qso10} that are members of LQGs and,
as in Clo13b, consider in particular, the range $1.1 \le \bar{z}_{LQG} \le
1.5$ then, from the one-sided Mann-Whitney test, we find a stronger shift ---
$0.103 \pm 0.015$ in the logarithm --- to smaller values of $\log_{10}L3000$
for the ultrastrong emitters (120 quasars) relative to the strong emitters
(463 quasars), at a level of significance given by a p-value of $2.024 \times
10^{-9}$. (Compare with the shift of $0.046 \pm 0.006$ obtained above for
{\sc dr7qso10} quasars in general.) Here too, there is no relative shift in
the distributions of $\log_{10}L3000$ for the two weaker intervals
weak\_upper and weak\_middle (Table~\ref{mw_summary_table}).

The LQGs are as defined in Clo13b, but here we are using only those members
that are also from {\sc dr7qso10}. Recall that {\sc dr7qso10} has the
criterion $sn\_med \ge 10.0$, which is needed here because, unlike Clo13b, we
are not making (and cannot make) comparisons with matched control samples.

If we consider the quasars from {\sc dr7qso10} that are not members of LQGs
then we find a shift to smaller values of $\log_{10}L3000$ of $0.042 \pm
0.006$ for the ultrastrong emitters relative to the strong emitters
(Table~\ref{mw_summary_table}), which is comparable to the shift ($0.046 \pm
0.006$) for the {\sc dr7qso10} quasars in general, since LQG members
constitute only a small fraction of the total ($\sim$ 11 per cent). Again,
there is no relative shift for weak\_upper and weak\_middle
(Table~\ref{mw_summary_table}).

We now briefly summarise the results so far. We find that for quasars in
general the highest Fe~{\sc II} emission, measured by $W2400$, tends to be
associated with the lowest intrinsic continuum fluxes, measured by $L3000$,
and expressed as $\log_{10}L3000$. We refer to this as the universal
dependence, because it applies to quasars in general. This tendency for
$W2400$ to increase as $\log_{10}L3000$ decreases is not strictly a
correlation (so not a Baldwin effect) as it appears to take effect only for
$W2400 \ga 25$~\AA. In accord with Clo13b, there appears to be a further,
differential, dependence for those quasars that are members of LQGs,
especially those with $1.1 \le z_{LQG} \le 1.5$, corresponding to a
more marked shift of the highest $W2400$ to lower values of $\log_{10}L3000$.

\begin{figure*}
\includegraphics[height=80mm]{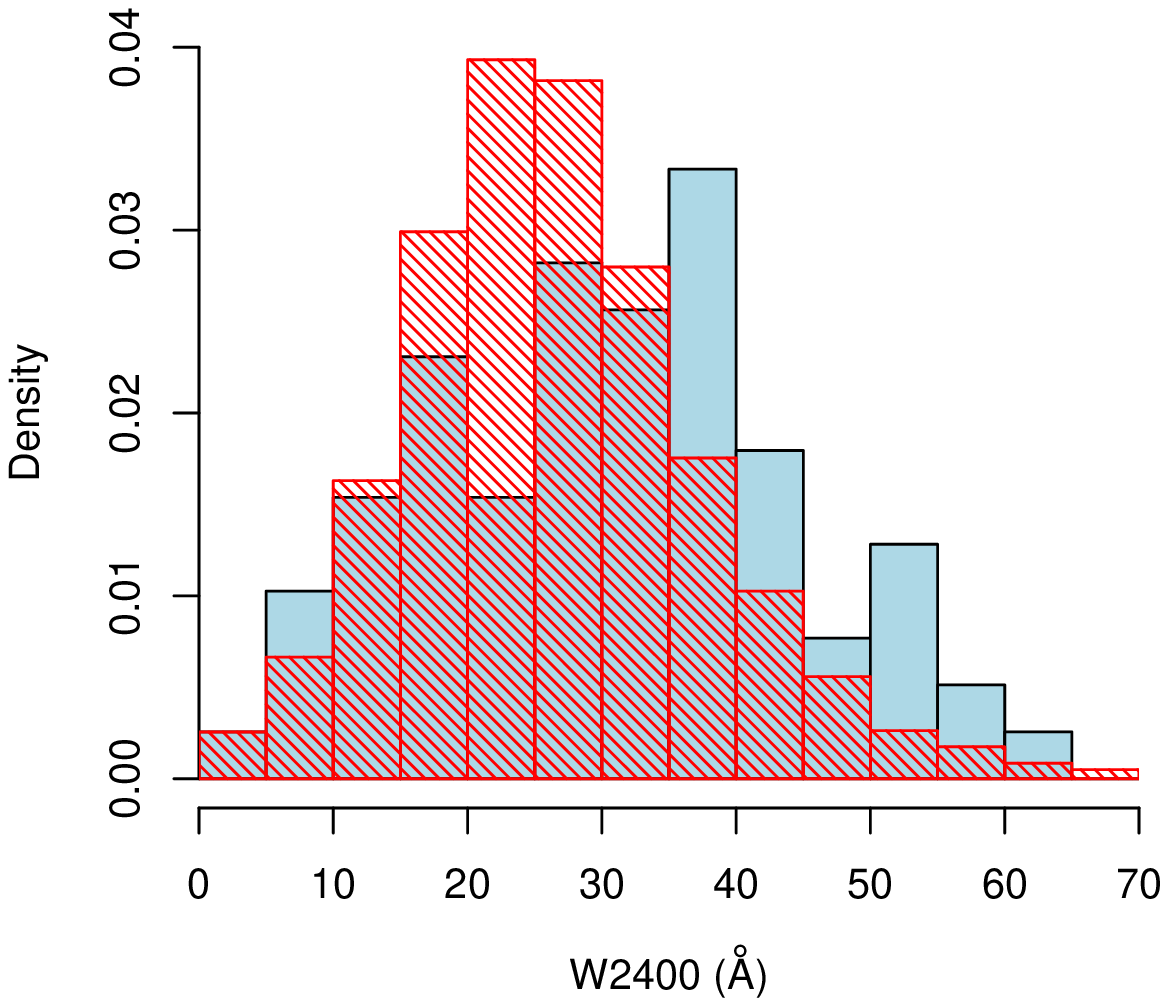}
\caption{The distributions of the rest-frame equivalent width, $W2400$, for
  the 80 quasars (from 83) of the sample {\sc sfqs10} (solid histogram, blue
  online) and for the 25700 quasars (from 25742) of the sample {\sc dr7qso10}
  (hatched histogram, red online). (Recall that for the 80 and the 25700 we
  can measure all of $W2400$, $fwhm2798$, and $L3000$.) Both are density
  histograms. The bin size is 5~\AA. The histograms have been restricted to
  $0 \le W2400 \le 70$~\AA\ for clarity. The universal dependence --- the
  tendency of the highest Fe~{\sc II} $W2400$ emission to be associated with
  the lowest intrinsic continuum luminosities --- is clearly seen as the
  relative preponderance of strong and ultrastrong emitters for the generally
  lower-luminosity sample {\sc sfqs10}.  }
\label{sfqs_w2400_sdssdr7_w2400_F1}
\end{figure*}

We can illustrate further the universal dependence in
Fig.~\ref{sfqs_w2400_sdssdr7_w2400_F1}, which shows the distributions of
$W2400$ of the samples {\sc sfqs10} and {\sc dr7qso10}, corresponding to
Fig.~\ref{LOG_L_3000_w2400_F1_c}. The sample {\sc sfqs10} is generally of
lower luminosity than {\sc dr7qso10}, and its relative preponderance of
strong and ultrastrong values of $W2400$ is very clear in the Figure.

\begin {table*}
\flushleft
\caption {A summary of the most important results from the Mann-Whitney tests
for the shift in $\log_{10}L3000$ according to the strength-category of
$W2400$. The columns are as follows.  (1)~Name of the sample.  (2)~Qualifier
on the sample (e.g.\ ``LQG members \ldots'').  (3)~The comparison being made ---
e.g.\ ultrastrong emitters compared with strong emitters.  (4)~The shift in
$\log_{10}L3000$ from the Hodges-Lehmann estimator together with an estimate
of the uncertainty.  (5)~The p-value from the one-sided Mann-Whitney test.
(6)~The sample sizes corresponding to column (3).  (7)~Notes.  Recall that
the strength-categories for $W2400$ are: ultrastrong with $W2400 \ge
45$~\AA; strong with $30 \le W2400 < 45$~\AA; weak with $W2400 <
30$~\AA. We sub-divide the weak category: weak\_upper with $20 \le W2400 <
30$~\AA; weak\_middle with $10 \le W2400 < 20$~\AA; weak\_lower with $W2400
< 10$~\AA.
}
\small \renewcommand \arraystretch {0.8}
\newdimen\padwidth
\setbox0=\hbox{\rm0}
\padwidth=0.3\wd0
\catcode`|=\active
\def|{\kern\padwidth}
\newdimen\digitwidth
\setbox0=\hbox{\rm0}
\digitwidth=0.7\wd0
\catcode`!=\active
\def!{\kern\digitwidth}
\begin {tabular} {llllllll}
\\
(1)               & (2)                             & (3)              & (4)                 & (5)                     & (6)           & (7)                    \\
Sample            & Qualifier                       & Comparison       & Shift in            & p-value                 & Sample sizes  & Notes                  \\
                  &                                 &                  & $\log_{10}L3000$    &                         &               &                        \\

\\ \hline
\\
{\sc dr7qso10}    & in general                      & ultrastrong cf.  & $0.046 \pm 0.006$   & $1.260 \times 10^{-12}$ & 1531 cf.      &                        \\
                  &                                 & strong           &                     &                         & 7086          &                        \\
\\
{\sc dr7qso10}    & in general                      & strong cf.       & $0.050 \pm 0.003$   & $< 2.2 \times 10^{-16}$ & 7086 cf.      &                        \\
                  &                                 & weak\_upper      &                     &                         & 9845          &                        \\
\\
{\sc dr7qso10}    & in general                      & weak\_upper cf.  & $0.000 \pm 0.004$   & $0.504$                 & 9845 cf.      &                        \\
                  &                                 & weak\_middle     &                     &                         & 5871          &                        \\
\\
{\sc dr7qso10}    & in general                      & ultrastrong cf.  & $0.096 \pm 0.006$   & $< 2.2 \times 10^{-16}$ & 1531 cf.      &                        \\
                  &                                 & weak\_upper      &                     &                         & 9845          &                        \\
\\ \hline
\\
{\sc sfqs10}      & in general                      & ultrastrong cf.  & $0.376 \pm 0.093$   & $3.545 \times 10^{-4}$  & 12 cf.        & No LQGs known          \\
                  &                                 & strong           &                     &                         & 30            &                        \\
\\
{\sc sfqs10}      & in general                      & strong cf.       & $0.107 \pm 0.106$   & $0.181$                 & 30 cf.        & No LQGs known          \\
                  &                                 & weak\_upper      &                     &                         & 17            &                        \\
\\
{\sc sfqs10}      & in general                      & weak\_upper cf.  & $-0.166 \pm 0.123$  & $0.892$                 & 17 cf.        & No LQGs known          \\
                  &                                 & weak\_middle     &                     &                         & 15            &                        \\
\\
{\sc sfqs10}      & in general                      & ultrastrong cf.  & $0.530 \pm 0.138$   & $0.003$                 & 12 cf.        & No LQGs known          \\
                  &                                 & weak\_upper      &                     &                         & 17            &                        \\
\\ \hline
\\
{\sc dr7qso10}    & LQG members                     & ultrastrong cf.  & $0.074 \pm 0.016$   & $1.613 \times 10^{-5}$  & 189 cf.       &                        \\
                  & all $\bar{z}_{LQG}$             & strong           &                     &                         & 790           &                        \\
\\
{\sc dr7qso10}    & LQG members                     & strong cf.       & $0.033 \pm 0.009$   & $6.070 \times 10^{-4}$  & 790 cf.       &                        \\
                  & all $\bar{z}_{LQG}$             & weak\_upper      &                     &                         & 1092          &                        \\
\\
{\sc dr7qso10}    & LQG members                     & weak\_upper cf.  & $-0.003 \pm 0.010$  & $0.5887$                & 1092 cf.      &                        \\
                  & all $\bar{z}_{LQG}$             & weak\_middle     &                     &                         & 657           &                        \\
\\
{\sc dr7qso10}    & LQG members                     & ultrastrong cf.  & $0.107 \pm 0.015$   & $1.118 \times 10^{-9}$  & 189 cf.       &                        \\
                  & all $\bar{z}_{LQG}$             & weak\_upper      &                     &                         & 1092          &                        \\
\\ \hline
\\
{\sc dr7qso10}    & LQG members                     & ultrastrong cf.  & $0.103 \pm 0.015$   & $2.024 \times 10^{-9}$  & 120 cf.       &                        \\
                  & $1.1 \le \bar{z}_{LQG} \le 1.5$ & strong           &                     &                         & 463           &                        \\
\\
{\sc dr7qso10}    & LQG members                     & strong cf.       & $0.034 \pm 0.010$   & $1.680 \times 10^{-3}$  & 463 cf.       &                        \\
                  & $1.1 \le \bar{z}_{LQG} \le 1.5$ & weak\_upper      &                     &                         & 638           &                        \\
\\
{\sc dr7qso10}    & LQG members                     & weak\_upper cf.  & $0.000 \pm 0.012$   & $0.5105$                & 638 cf.       &                        \\
                  & $1.1 \le \bar{z}_{LQG} \le 1.5$ & weak\_middle     &                     &                         & 386           &                        \\
\\
{\sc dr7qso10}    & LQG members                     & ultrastrong cf.  & $0.133 \pm 0.016$   & $9.088 \times 10^{-14}$ & 120 cf.       &                        \\
                  & $1.1 \le \bar{z}_{LQG} \le 1.5$ & weak\_upper      &                     &                         & 638           &                        \\
\\ \hline
\\
{\sc dr7qso10}    & not LQG members                 & ultrastrong cf.  & $0.042 \pm 0.006$   & $1.799 \times 10^{-9}$  & 1342 cf.      &                        \\
                  &                                 & strong           &                     &                         & 6296          &                        \\
\\
{\sc dr7qso10}    & not LQG members                 & strong cf.       & $0.052 \pm 0.003$   & $< 2.2 \times 10^{-16}$ & 6296 cf.      &                        \\
                  &                                 & weak\_upper      &                     &                         & 8753          &                        \\
\\
{\sc dr7qso10}    & not LQG members                 & weak\_upper cf.  & $0.000 \pm 0.004$   & $0.4782$                & 8753 cf.      &                        \\
                  &                                 & weak\_middle     &                     &                         & 5214          &                        \\
\\
{\sc dr7qso10}    & not LQG members                 & ultrastrong cf.  & $0.094 \pm 0.006$   & $< 2.2 \times 10^{-16}$ & 1342 cf.      &                        \\
                  &                                 & weak\_upper      &                     &                         & 8753          &                        \\
\\ \hline
\\
\end {tabular}
\\
\label{mw_summary_table}
\end {table*}

\subsubsection{BALs in {\sc sfqs10}}
\label{subsubsecBALsfqs10}

A higher rate of BALs for faint quasars compared with brighter quasars could
conceivably introduce some component of the shift to stronger $W2400$
values. We have examined the spectra of the 55 {\sc sfqs10} quasars with
$\log_{10}L3000 \le 45.4$ that contribute to
Fig.~\ref{sfqs_w2400_sdssdr7_w2400_F1} and Fig.~\ref{LOG_L_3000_w2400_F1_c}
to judge whether there is any indication of an unusually high frequency of
BALs. Recall that the {\sc sfqs10} quasars are generally lower-luminosity
quasars --- see Fig.~\ref{LOG_L_3000_w2400_F1_c}.

We find no occurrences of the relatively rare, low-ionisation Mg~{\sc II}
troughs in the spectra of these 55 quasars.
However, high-ionisation C~{\sc IV}~$\lambda 1549$ troughs would not be
detectable in the spectra of those {\sc sfqs10} quasars that have $z \la
1.5$. For C~{\sc IV} BALs we are therefore limited to considering only those
{\sc sfqs10} spectra for which $z \ga 1.5$: 13 from the 55 have $z \ge
1.50$. Of these 13, one, at $z = 1.64$, appears very likely to be a BAL; it
has $W2400 = 43.2$~\AA, which is in the strong category, but close to being
in the ultrastrong. Two of the 13 quasars have higher values of $W2400$, both
in the ultrastrong category: one is clearly not a BAL; the other is almost
certainly not a BAL, but its redshift is close to the limit for
detection. Consequently, although the opportunities here to recognise BALs
are evidently few, there is at least no reason to suppose that for $z \ge
1.50$ the frequency and properties of BAL quasars are different from those at
brighter continuum luminosities. We shall assume that the same applies also
to $z \la 1.5$.

\subsection{${\bf W2400}$ and line-widths} \label{subsecLineWidths}

We have previously suspected that ultrastrong UV Fe~{\sc II} emission tends to
be associated with quasars that have broad emission lines that are unusually
narrow.

\begin{figure*}
\includegraphics[height=80mm]{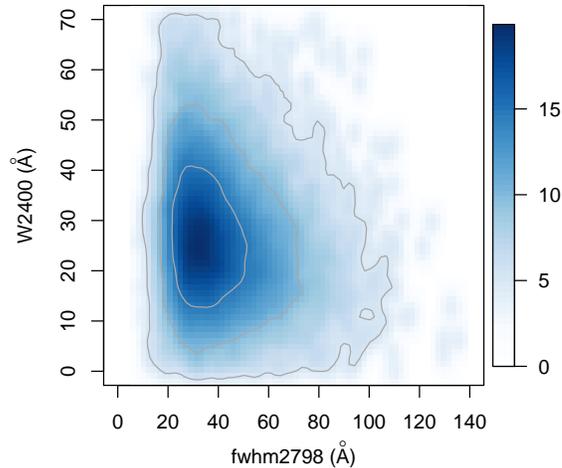}
\caption{A plot of $W2400$ against $fwhm2798$, the FWHM of the Mg~{\sc II}
$\lambda 2798$, emission for the sample {\sc dr7qso10}. The shading
indicates the kernel-smoothed densities of points in a $64 \times 64$ grid,
because the number of points, 25700, is too high for an ordinary
scatterplot to be a useful illustration. The plot has been restricted to
$0 \le fwhm2798 \le 140$~\AA\ and $0 \le W2400 \le 70$~\AA\ for clarity.
Note that the highest Fe~{\sc II} emitters tend to favour relatively
narrow Mg~{\sc II} emission lines.
}
\label{fwhm2798_w2400_F1_c}
\end{figure*}

We can therefore perhaps gain some understanding of the universal dependence
of the $W2400$ emission on $L3000$ and the differential dependence for the
members of LQGs by considering Fig.~\ref{fwhm2798_w2400_F1_c}. It plots
$W2400$ against $fwhm2798$, the FWHM of the Mg~{\sc II} $\lambda 2798$
emission (calculated as described in section~\ref{secMeasuring}), for the
sample {\sc dr7qso10}. Recall that {\sc dr7qso10} specifies: $1.0 \le z \le
1.8$, $i \le 19.1$, and $sn\_med \ge 10.0$; it has 25742 quasars. We can
measure all of $W2400$, $fwhm2798$, and $L3000$ for 25700 of the {\sc
  dr7qso10} quasars. The number of points is too high for an ordinary
scatterplot to be a useful illustration and so the plot shows instead the
kernel-smoothed densities of points in a $64 \times 64$ grid. Note that the
highest Fe~{\sc II} emitters tend to favour relatively narrow Mg~{\sc II}
emission lines. It is not an exclusive relation however: high Fe~{\sc II}
emission does not guarantee narrow Mg~{\sc II}, and narrow Mg~{\sc II}
emission does not guarantee high Fe~{\sc II} emission.

\begin{figure*}
\includegraphics[height=80mm]{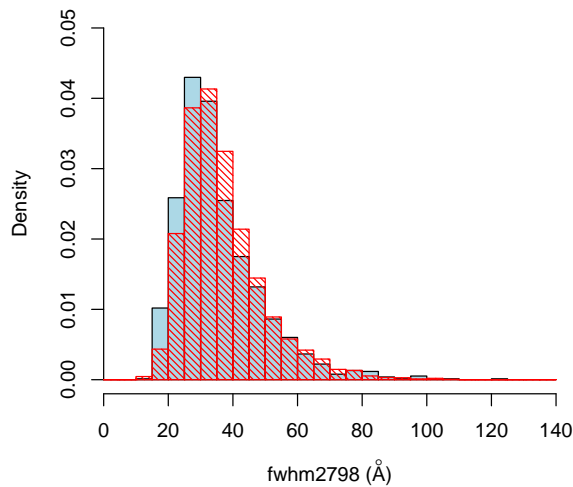}
\caption{The distributions of the FWHM, $fwhm2798$, of the Mg~{\sc II}
emission line for the 1531 ultrastrong $W2400$ emitters of the {\sc
dr7qso10} sample (solid histogram, blue online) and for the 7086 strong
$W2400$ emitters (hatched histogram, red online). Both are density
histograms. The bin size is 5~\AA. The histograms have been restricted to
$0 \le fwhm2798 \le 140$~\AA\ for clarity. The tendency of the highest
Fe~{\sc II} emission to favour relatively narrow Mg~{\sc II} emission is
clearly seen.}
\label{fwhm2798_F1}
\end{figure*}

This tendency of the highest Fe~{\sc II} emission favouring relatively narrow
Mg~{\sc II} emission is further illustrated in Fig.~\ref{fwhm2798_F1}, which
shows, for the {\sc dr7qso10} sample, the distributions of $fwhm2798$ for the
1531 ultrastrong emitters (solid histogram, blue online) and for the 7086
strong emitters (hatched histogram, red online). Of course, the boundary
between strong and ultrastrong is somewhat arbitrary, but the difference in
the distributions is nevertheless very clear.

A one-sided Mann-Whitney test indicates that there is a relative shift to
smaller values of the $fwhm2798$ distribution for the ultrastrong emitters at
a level of significance given by a p-value of $1.522 \times 10^{-9}$. The
median shift is estimated as $1.89 \pm 0.28$~\AA.
A similar result is obtained if instead we use the FWHM\_MGII parameter
(whole profile --- i.e.\ broad $+$ narrow) from \citet{Shen2011}: p-value of
$3.504 \times 10^{-16}$ and median shift of $2.40 \pm 0.26$~\AA\ (for 1530
ultrastrong emitters and 7065 strong emitters here, since FWHM\_MGII is
available only for 25613 of the 25700 quasars)\footnote{We find a tendency
  for stronger UV Fe~{\sc II} to be associated with narrower Mg~{\sc
    II}. Possible trends for the UV Fe~{\sc II} EW and Mg~{\sc II} FWHM
  appear not to be mentioned in \citet{Shen2011}. However, their Fig.~13
  includes a small contour plot (leftmost column, fourth from top), with
  linear-interval contours but logarithmic axes, that might, uncertainly,
  suggest the opposite tendency of stronger Fe~{\sc II} being associated with
  broader Mg~{\sc} II. We note the following: their plot is for all quasars
  so no s/n threshold has been applied, and the plot is therefore not
  strictly suitable for comparison; a further small contour plot in their
  Fig.~13 (leftmost column, bottom) indicates that their broadest Mg~{\sc II}
  lines tend to correspond to spectra of low s/n; their EW for Fe~{\sc II} is
  presumably their UV~Fe~EW measure for 2200-3090~\AA\ ($EW\_FE\_MGII$),
  which is different from our $W2400$ measure; there is no statement of which
  of their measures of Mg~{\sc II} FWHM has been used; we mentioned in an
  earlier footnote our suspicion that in their measures involving subtraction
  of a narrow component (at least), residual (after subtraction of the iron
  template) neighbouring Fe~{\sc II} can be wrongly attributed to broad
  Mg~{\sc II}; our method of measuring the FWHM of the Mg~{\sc II} was
  intended to avoid neighbouring Fe~{\sc II}.}.

The tendency is not detected with the {\sc sfqs10} sample (the 80 from 83 for
which we can measure all of $W2400$, $fwhm2798$, and $L3000$) by applying the
Mann-Whitney test to the ultrastrong (12 quasars) and strong emitters (30
quasars), and so it is presumably less marked than the effect for
$\log_{10}L3000$. Nevertheless, we can note the consistency of this tendency
with the properties of the {\sc sfqs10} sample: 23 of the 42 quasars (55 per
cent) that are ultrastrong or strong emitters have narrow $fwhm2798 <
35.0$~\AA, whereas only 13 of the 38 quasars (34 per cent) that are weak
emitters have $fwhm2798 < 35.0$~\AA.

Although we do not show it here, the tendency for $W2798$ is different:
higher $W2798$ tends to favour broader $fwhm2798$, as found also by
\citet{Shen2011}. If we assume that the Fe~{\sc II} emission and the Mg~{\sc
  II} line arise from the same locations within the BLR then, presumably, any
hypothetical orientation effects of the flattened BLR would be identical.  In
that case, we can suspect a direct connection between narrow $fwhm2798$ and
enhanced $W2400$. Narrow (BLR) Mg~{\sc II} line emission presumably
corresponds to narrow (BLR) Ly$\alpha$ emission. (The widths of Mg~{\sc II}
and Ly$\alpha$ are presumably both determined by the mass of the central BH,
but are unlikely to be identical because of stratification, Ly$\alpha$
absorption, Ly$\alpha$ blending with N~{\sc V} $\lambda 1240$, and other
factors.) We have few opportunities to test such a correspondence because
Ly$\alpha$ will be in the ultraviolet region when Mg~{\sc II} and Fe~{\sc II}
are observed in the optical region. However, from our own data, we do have
GALEX ultraviolet spectra for a few of the quasars for which we have optical
MMT / Hectospec spectra (see Appendix~\ref{secAppendixMMTHectospecGALEX}).
In Fig.~\ref{w1216_galex_example_spectra} we show the GALEX ultraviolet
spectra, converted to rest-frame wavelengths, for the ultrastrong emitters
qso412 and qso425, for which we previously showed optical spectra, converted
to rest-frame wavelengths, in Fig.~\ref{w2400_example_spectra}. Clearly,
qso412 has broad Mg~{\sc II} line emission, and qso425 has narrow Mg~{\sc II}
line emission. Clearly, Ly$\alpha$ is correspondingly broad for qso412 and
correspondingly narrow for qso425. For Ly$\alpha$ the ratio of FWHMs,
qso412~:~qso425, is 1.6 (from 24.4~\AA\ and 15.7~\AA), and for Mg~{\sc II} it
is 2.1 (from 50.6~\AA\ and 24.2~\AA). See
Appendix~\ref{secAppendixMMTHectospecGALEX} for further details. Also, note
again what we emphasised above: the {\it tendency} of the highest Fe~{\sc II}
emitters favouring relatively narrow Mg~{\sc II} emission does not exclude
the occurrence of quasars such as qso412 with ultrastrong $W2400$ and broad
Mg~{\sc II}.

\begin{figure*}
\includegraphics[height=80mm]{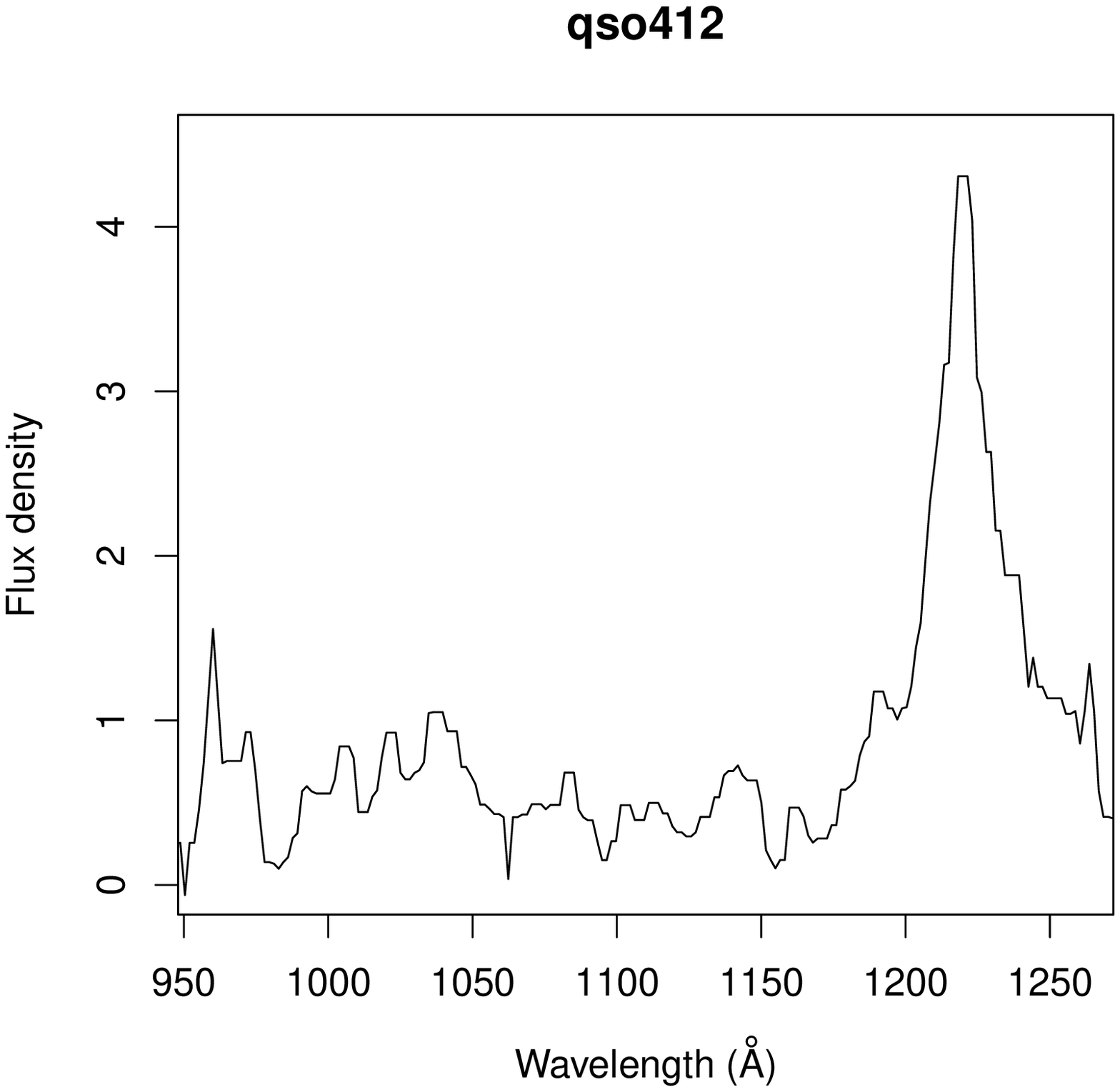}
\includegraphics[height=80mm]{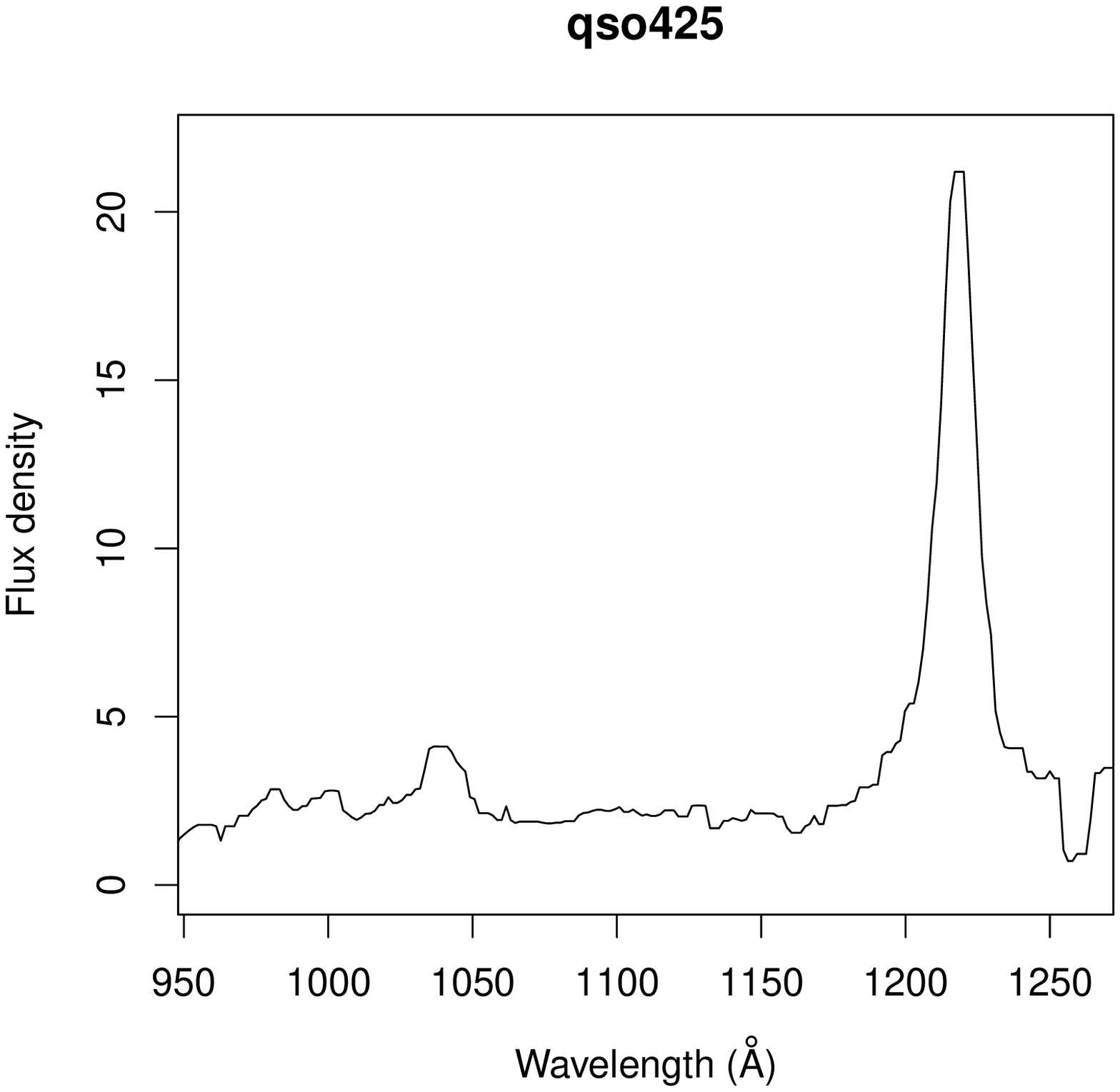}

\caption{
Rest-frame (ultraviolet) spectra of two ultrastrong UV Fe~{\sc II}-emitting
quasars, qso412 ($z = 1.156$) and qso425 ($z = 1.230$), from our GALEX
spectra (Appendix~\ref{secAppendixMMTHectospecGALEX}).  The rest-frame
optical spectrum (Fig.~\ref{w2400_example_spectra}) shows broad Mg~{\sc II}
$\lambda 2798$ line emission for qso412, and narrow for qso425. Here it can
be seen that the Ly$\alpha$ $\lambda 1216$ line emission is correspondingly
broad for qso412 and narrow for qso425. The spectra have been smoothed with a
5-pixel median filter. The flux densities ($f_\lambda$) are in units of
$10^{-16}$~erg~s$^{-1}$~cm$^{-2}$~\AA$^{-1}$. The wavelength range has been
restricted to $\sim$~950--1250~\AA\ for clarity. See
Table~\ref{mmt_hecto_quasars} for further details of these two quasars.
}
\label{w1216_galex_example_spectra}
\end{figure*}

For a given Ly$\alpha$ flux, a narrow line will concentrate relatively more
flux than a broad line in the region Ly$\alpha$ $\lambda 1216$ $\pm
3$~\AA\ \citep{Johansson1984, Sigut1998} that is most important for
Ly$\alpha$ fluorescence of the UV Fe~{\sc II}. We suggest, therefore, that
Ly$\alpha$ fluorescence might be in large part responsible for both the
universal dependence of $W2400$ on $L3000$ and the differential dependence
for LQG members.

\begin{figure*}
\includegraphics[height=80mm]{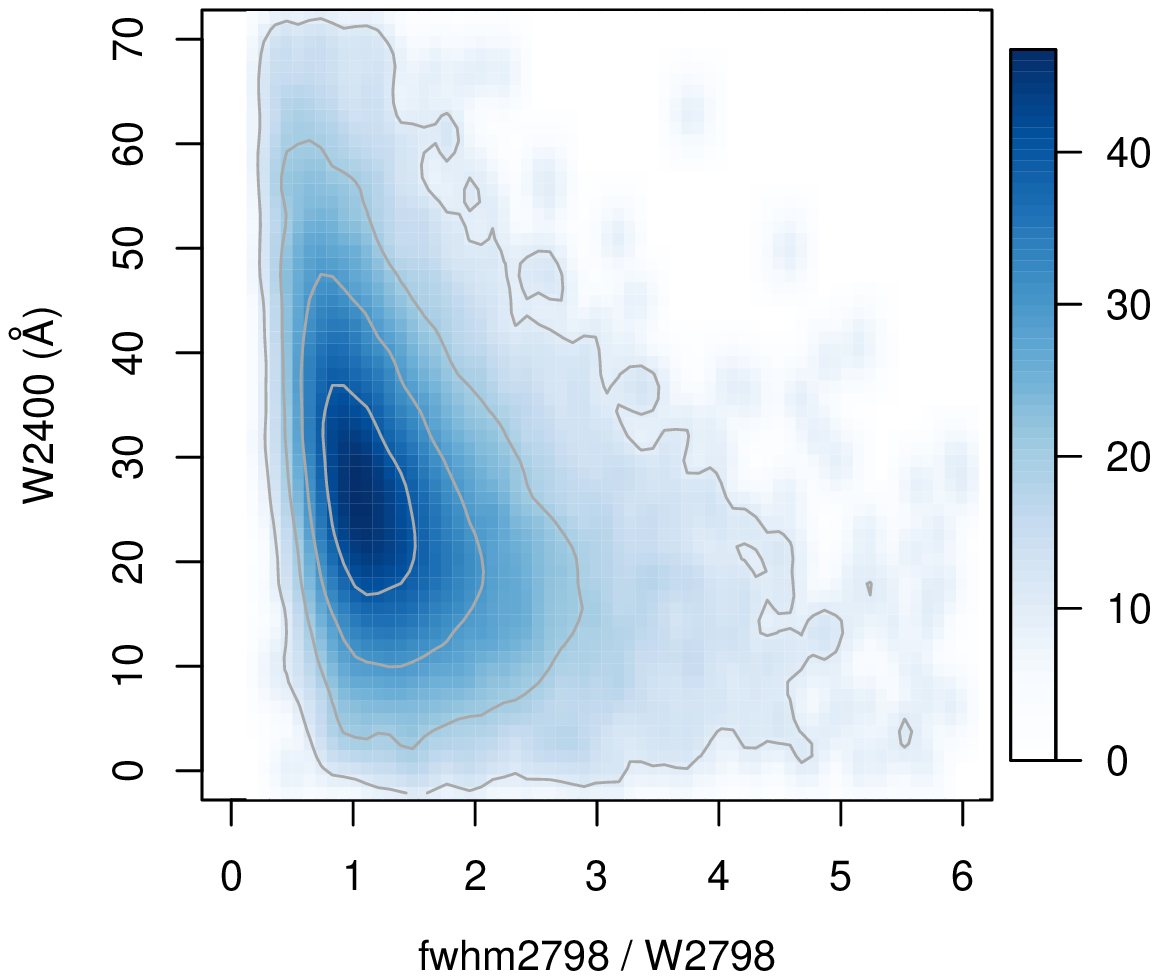}
\caption{A plot of $W2400$ against $fwhm2798 / W2798$, the FWHM of the
Mg~{\sc II} $\lambda 2798$ emission divided by the equivalent width, for
the sample {\sc dr7qso10}. The shading indicates the kernel-smoothed
densities of points in a $64 \times 64$ grid, because the number of points,
25700, is too high for an ordinary scatterplot to be a useful illustration.
The plot has been restricted to $0 \le fwhm2798 / w2798 \le 6.0$ and
$0 \le W2400 \le 70$~\AA\ for clarity. Note that increasing $W2400$ emission
seems to be associated with decreasing values of the ratio $fwhm2798 / W2798$.
}
\label{fwhm2798_w2400_F1_f}
\end{figure*}

We mentioned previously that, for Ly$\alpha$ fluorescence, the ratio of FWHM
to equivalent width of the Ly$\alpha$ would be a useful central-concentration
parameter for associating with the Fe~{\sc II} emission. For the spectral
coverage of our quasar data we cannot routinely calculate this ratio for
Ly$\alpha$, which then falls in the ultraviolet region. Perhaps, however,
Mg~{\sc II} can be used as at least a rough guide to the ratio for
Ly$\alpha$. In Fig.~\ref{fwhm2798_w2400_F1_f}, we plot $W2400$ against
$fwhm2798 / W2798$ for {\sc dr7qso10}. Again, the number of points (25700) is
too high for an ordinary scatterplot to be a useful illustration and the plot
shows instead the kernel-smoothed densities of points in a $64 \times 64$
grid. Note that increasing $W2400$ emission seems to be associated with
decreasing values of the ratio $fwhm2798 / W2798$. That is, increasingly
strong UV Fe~{\sc II} emission does seem to be associated with increasingly
centrally-concentrated Mg~{\sc II} emission at least. Note again, of course,
that the Mg~{\sc II} emission can be affected by Fe~{\sc II} emission,
especially on the blueward side.

\subsection{Interpretation} \label{subsecInterpretation}

\begin{figure*}
\includegraphics[height=80mm]{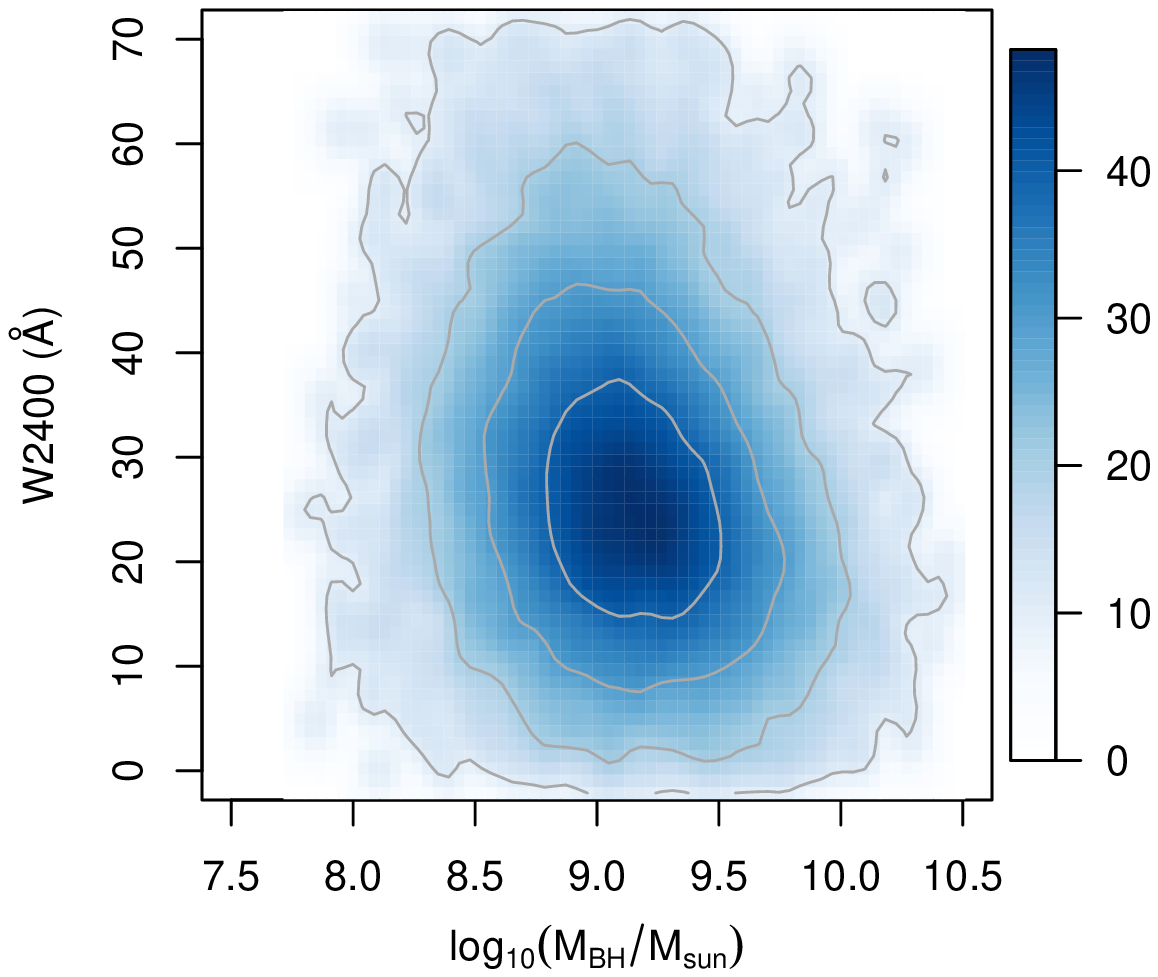}
\caption{A plot of $W2400$ against $\log_{10}(M_{\rm BH}/M_\odot)$, where
$M_{\rm BH}$ is the adopted ``fiducial'' virial BH mass from \citet{Shen2011},
for the sample {\sc dr7qso10}. The shading indicates the kernel-smoothed
densities of points in a $64 \times 64$ grid, because the number of points,
25670, is too high for an ordinary scatterplot to be a useful
illustration. The plot has been restricted to
$7.5 \le \log_{10}(M_{\rm BH}/M_\odot) \le 10.5$ and $0 \le W2400 \le 70$~\AA\
for clarity. Note that increasing $W2400$ emission seems to be associated
with decreasing values of $M_{\rm BH}$. 
}
\label{log10MBH_w2400_F1_c}
\end{figure*}

We might expect narrow emission lines to have some correspondence with
relatively less massive central BHs --- either to AGN that intrinsically have
low BH masses or to younger quasars that are still growing their central BHs
and have not yet reached their mature state \citep[e.g.][]{Mathur2000}. In
Fig.~\ref{log10MBH_w2400_F1_c}, we plot $W2400$ against $\log_{10}(M_{\rm
  BH}/M_\odot)$ for {\sc dr7qso10}, where $M_{\rm BH}$ is the adopted
``fiducial'' virial BH mass from \citet{Shen2011}. $M_{\rm BH}$ is actually
determined by a FWHM, acting as a proxy for virial velocity, and by a
continuum luminosity, acting as a proxy for BLR radius. Again, the number of
points (25670 of the previous 25700 have a measurement of $M_{\rm BH}$) is
too high for an ordinary scatterplot to be a useful illustration and the plot
shows instead the kernel-smoothed densities of points in a $64 \times 64$
grid. From Shen et al.\ typical errors in the BH masses, propagated from the
measurement uncertainties in the continuum luminosities and the FWHMs, are
$\sim$~0.05--0.2~dex, but the additional statistical uncertainty in the
calibration of virial masses is $\sim$~0.3--0.4~dex. Nevertheless, there does
appear to be a trend in Fig.~\ref{log10MBH_w2400_F1_c}, with average
ultrastrong emitters having smaller BH masses than average weak emitters by a
factor of $\sim 2$. Clearly, the ultrastrong emitters are not in the
low-BH-mass category ($\la 10^8M_\odot$) of AGN, and we might instead expect
that they are still growing their BHs. In that case, increasing $W2400$
emission would have an association with increasingly young quasars. We
further suggest that this possible interpretation in terms of younger quasars
helps to clarify the nature of LQGs. LQGs --- regions with an overdensity of
quasars --- would then be interpreted quite naturally as regions that contain
a higher proportion of young quasars than field regions.

\section{Further notes} \label{secFurtherNotes}

In this section we make some further notes on the content of the paper, some
of which have a slightly cautionary nature because of the possibilities of
small systematic biases in the measurements and because of the difficulty of
clarifying further the possible implications of the data.

\begin{figure*}
\includegraphics[height=80mm]{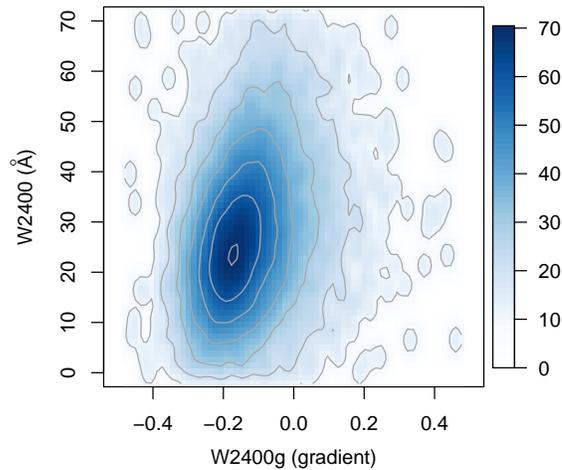}
\caption{A plot of $W2400$ against $W2400g$, a measure of the gradient,
represented as a colour, of the line taken as the continuum in the
\citet{Weymann1991} definition of the equivalent width $W2400$,
for the sample {\sc dr7qso10}. The shading indicates the kernel-smoothed
densities of points in a $64 \times 64$ grid, because the number of points,
25700, is too high for an ordinary scatterplot to be a useful
illustration. The plot has been restricted to $-0.5 \le W2400g \le
0.5$ and $0 \le W2400 \le 70$~\AA\ for clarity. Note the clear
correlation, which might be exaggerated by a small bias in $W2400$ for
red quasars with respect to blue, but which cannot be attributed to it.
}
\label{w2400g_w2400_F1_c}
\end{figure*}

First of all, we note that there is a correlation of $W2400$ with $W2400g$
(Fig.~\ref{w2400g_w2400_F1_c}). Recall from section~\ref{secMeasuring} that
$W2400g$ is a measure of the gradient of the line taken as the continuum in
the \citet{Weymann1991} definition of the equivalent width $W2400$. It is
represented as a colour. A similar correlation results if instead of $W2400g$
we use the similarly-defined continuum colour
$-2.5\log_{10}[(L2200/2200)/(L3000/3000)]$, which corresponds to a larger
interval of wavelength than $W2400g$. The correlations are presumably
counterparts of correlations with spectral index.

In connection with Fig.~\ref{w2400_example_spectra}, in
section~\ref{secMeasuring}, we discussed briefly the effect of artificially
changing the curvature of the spectrum of qso412 to resemble that of qso425.
It reduced $W2400$ from 55.5~\AA\ to 52.1~\AA, the difference of
3.4~\AA\ being comparable to the indicative measurement errors.  Similarly,
for {\sc dr7qso10}, we find that $W2400$ for the red quasars ($W2400g > 0$,
1292 from 25700 --- so only $\sim$ 5 per cent of the total) in
Fig.~\ref{w2400g_w2400_F1_c} could be biased with respect to $W2400$ for the
blue quasars by $\sim 3$~\AA. If so, the correlation might then be
exaggerated by a small bias in $W2400$ for red quasars with respect to blue
but cannot be attributed to it. However, by artificially changing the
curvature of the spectra we are clearly also changing the quasars (from red
to blue): it may be that the measurements of $W2400$ for the red quasars are
mostly legitimate, and not subject to any such small bias.

The $1/\lambda$ function by which we artificially change the curvature of a
spectrum (here and as mentioned in section~\ref{secMeasuring}) is this:
first, we multiply by a function of $1/\lambda$ such that $f_{\lambda}(3000)$
is unchanged and $f_{\lambda}(1900)$ is doubled; then we scale the spectrum
such that the median flux density across 2255‒-2650~\AA\ is unchanged.

We mentioned previously, in a footnote, that we have reasons to prefer our
measure $W2400$ to the UV~EW~Fe ($EW\_FE\_MGII$) measure from
template-fitting of \citet{Shen2011}. Nevertheless, we note that their
measure also correlates with $W2400g$ and with the continuum colour
$-2.5\log_{10}[(L2200/2200)/(L3000/3000)]$, although the trends with their
measure have more complexity or internal structure.

\begin{figure*}
\includegraphics[height=80mm]{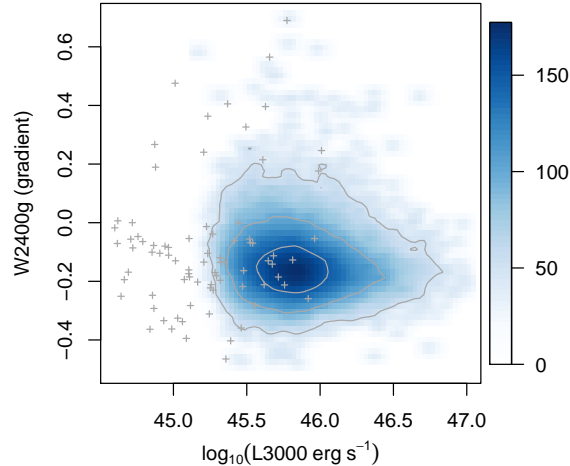}
\caption{A plot of $W2400g$ against $\log_{10}L3000$ for the sample {\sc
 dr7qso10}. $L3000$ is the intrinsic continuum flux, in units of
erg~s$^{-1}$, at 3000~\AA\ in the rest frame (see the text for its
definition). $W2400g$ is a measure of the gradient, represented as a
colour, of the line taken as the continuum in the \citet{Weymann1991}
definition of the equivalent width $W2400$. The shading indicates the
kernel-smoothed densities of points in a $64 \times 64$ grid, because the
number of points, 25700, is too high for an ordinary scatterplot to be a
useful illustration. Linear contours for the densities are also shown. The
plot has been restricted to $44.6 \le log_{10}L3000 \le 47.0$ and $-0.5 \le
W2400g \le 0.7$ for clarity. The plot also shows, as points
(crosses), $W2400g$ against $\log_{10}L3000$, for the sample {\sc sfqs10}
(80 quasars). Note that $W2400g$ appears to be only slightly dependent on
$\log_{10}L3000$.
}
\label{LOG_L_3000_w2400g_F1_c}
\end{figure*}

This dependence of $W2400$ on $W2400g$ and on the similar measure of
continuum colour $-2.5\log_{10}[(L2200/2200)/(L3000/3000)]$ (and similarly
for the Shen et al. UV~EW~Fe measure) thus appears to be a small dependence
of the UV Fe on continuum colour. Note that it does not appear to arise from
the dependence of $W2400$ on $L3000$ that we find
(section~\ref{subsecFainterW2400}): $W2400g$ appears to be only slightly
dependent on $L3000$ --- see Fig.~\ref{LOG_L_3000_w2400g_F1_c}, which plots
$W2400g$ against $\log_{10}L3000$ as kernel-smoothed densities for the sample
{\sc dr7qso10} and as points for the sample {\sc sfqs10}. For illustration,
the Mann-Whitney test applied to the {\sc dr7qso10} data, shows that there is
a significant (p-value $< 2.2 \times 10^{-16}$) but small dependence of
$W2400g$ on $\log_{10}L3000$ --— corresponding to a colour shift of only
$\Delta(W2400g) \sim 0.025 \pm 0.002$ for the interval $45.4 \le
\log_{10}L3000 < 45.5$ compared with the interval $45.9 \le \log_{10}L3000
\le 46.0$.
Of course our finding that stronger $W2400$ tends to be associated with
fainter $L3000$ combined with the correlation of $W2400$ and $W2400g$ means
that there must be some tendency, even if small, for $W2400g$ to be redder
for fainter $L3000$.

We note that the modelling by \citet{Verner2003}, \citet{Verner2004} and
\citet{Bruhweiler2008} appears to take spectral index as a fixed quantity
rather than a variable. \citet{Bruhweiler2008} do, however, mention that the
shape of the spectral energy distribution would be a consideration in future
work, although we have not found any subsequent references in which it is
considered.

\section{Conclusions} \label{secConclusions}

The conclusions of this paper are the following.

\begin{enumerate}

\item We find that for $W2400 \ga 25$~\AA\ there is a {\it universal\/}
  (i.e.\ for quasars in general) strengthening of $W2400$ with decreasing
  intrinsic luminosity, $L3000$.
\item In conjunction with the work presented by \citet{Clowes2013b}, we find
  that there is a further, {\it differential,} strengthening of $W2400$ with
  decreasing $L3000$ for those quasars that are members of Large Quasar
  Groups, or LQGs.
\item We find that increasingly strong $W2400$ tends to be associated with
  decreasing FWHM of the neighbouring Mg~{\sc II} $\lambda 2798$ broad
  emission line.
\item On the assumption that this trend that $W2400$ increases as the FWHM of
  Mg~{\sc II} decreases is true also for the FWHM of the Ly$\alpha$ emission
  line, we suggest that this dependence of $W2400$ on intrinsic luminosity
  arises from Ly$\alpha$ fluorescence. We make this suggestion because: the
  wavelength region $1216 \pm 3$~\AA\ is important for Ly$\alpha$
  fluorescence of the Fe~{\sc II} emission; for a given Ly$\alpha$ equivalent
  width, narrower Ly$\alpha$ emission would enhance the flux in this region,
  and hence fluorescence of the Fe~{\sc II}, relative to broader Ly$\alpha$
  emission.
\item We find that stronger $W2400$ tends to be associated with smaller
  virial estimates of the mass of the central black hole, by a factor $\sim
  2$ between the ultrastrong emitters (smaller black holes) and the weak. The
  effect for $W2400$ can then be associated with a range of masses for the
  central black holes. Stronger $W2400$ emission would correspond to smaller
  black holes that are still growing. They will be in fainter quasars, with
  relatively narrow broad-emission lines, leading to the enhanced Ly$\alpha$
  fluorescence of the UV Fe~{\sc II} emission. The differential effect for
  LQGs might then arise from a different mass distribution of the black
  holes, corresponding, plausibly, to an age distribution that emphasises
  younger quasars in the LQG environments.
\end{enumerate}

\section{Acknowledgments}

We wish to thank: Linhua Jiang for providing us with the SFQS spectra in
digital form; and Martin Gaskell for very informative discussions on the
characteristics of the BLRs in quasars.
 
The anonymous referee is thanked for a thoughtful and constructive review of
the paper.

We also wish to thank the {\sc R} Foundation for Statistical Computing for
the {\sc R} software.

LEC received partial support from the Center of Excellence in Astrophysics
and Associated Technologies (PFB 06), and from a CONICYT Anillo project (ACT
1122).

SR was in receipt of a CONICYT PhD studentship while much of this work was in
progress.

This research has used the SDSS DR7QSO catalogue \citep{Schneider2010}.

Funding for the SDSS and SDSS-II has been provided by the Alfred P. Sloan
Foundation, the Participating Institutions, the National Science
Foundation, the U.S. Department of Energy, the National Aeronautics and
Space Administration, the Japanese Monbukagakusho, the Max Planck
Society, and the Higher Education Funding Council for England. The SDSS
Web Site is http://www.sdss.org/.

The SDSS is managed by the Astrophysical Research Consortium for the
Participating Institutions. The Participating Institutions are the American
Museum of Natural History, Astrophysical Institute Potsdam, University of
Basel, University of Cambridge, Case Western Reserve University, University
of Chicago, Drexel University, Fermilab, the Institute for Advanced Study,
the Japan Participation Group, Johns Hopkins University, the Joint Institute
for Nuclear Astrophysics, the Kavli Institute for Particle Astrophysics and
Cosmology, the Korean Scientist Group, the Chinese Academy of Sciences
(LAMOST), Los Alamos National Laboratory, the Max-Planck-Institute for
Astronomy (MPIA), the Max-Planck-Institute for Astrophysics (MPA), New Mexico
State University, Ohio State University, University of Pittsburgh, University
of Portsmouth, Princeton University, the United States Naval Observatory, and
the University of Washington.

\appendix

\section{MMT / Hectospec and GALEX data} \label{secAppendixMMTHectospecGALEX}

In this paper we have made a little use of our own MMT / Hectospec spectra
and GALEX spectra. We give here some more details, especially for the MMT /
Hectospec spectra. Table~\ref{mmt_hecto_quasars} gives our measurements of
$W2400$ and other parameters from these MMT / Hectospec spectra, primarily
for the redshift range $1.0 \le z \le 1.8$, but, for completeness, with some
information for the quasars outside this range included too.

The MMT / Hectospec spectra were obtained in 1.6~deg$^2$ from two adjacent
$0.8~{\rm deg}^2$ Hectospec fields. The fields intersect the LQGs U1.11,
U1.28, and U1.54 \citep{Clowes2012, Clowes1991}, with the last of these being
the ``doubtful LQG'' (a previously published LQG, but one of marginal
significance in our catalogue). Quasars in the redshift range $1.0 \le z \le
1.8$ are of particular interest to us, because this is the range
corresponding to our main DR7QSO catalogue of LQGs.

The RA, Dec (2000) field centres were approximately: 10:48:40.8, $+$05:24:36
($162.17\degr, 5.41\degr$) and 10:50:04.8, $+$04:31:12 ($162.52\degr,
4.52\degr$), but some small adjustments were made during the observations
across two months. Spectra were obtained during February and April 2010,
covering 3650--9200~\AA\ at $\sim$~6~\AA\ resolution, with integration times
typically 5400~s but with some variation. As mentioned previously, we used
the wavelength range 3900--8200~\AA\ in the measurement of $W2400$ and other
parameters from our MMT / Hectospec spectra. Also as mentioned previously,
for our MMT / Hectospec spectra only, the atmospheric A-band at
$\sim$~7600~\AA\ is prominent and we have interpolated across it. For the
redshift range $1.0 \le z \le 1.8$ the measurements of $W2400$ do not contain
the interpolation. For two of these quasars, noted in
Table~\ref{mmt_hecto_quasars}, the Mg~{\sc II} emission does contain the
interpolation and the measurements of $fwhm2798$ could be affected a
little. For three quasars outside the range $1.0 \le z \le 1.8$ the
measurements of $W2400$ do contain the interpolation, and these too have been
noted in Table~\ref{mmt_hecto_quasars}.

The MMT / Hectospec observations of the quasars were ``fibre-filling''
observations --- using spare fibres after the objects for the main programme
(on galaxies) had all been allocated fibres. The quasar candidates that were
selected for observation were drawn from a list that had been prepared
previously using the SDSS DR1 photometric quasar catalogue of
\citet{Richards2004}. All of the candidates in the list were selected to have
$18.0 \le r \le 21.2$ (the catalogue itself was defined by $g \le 21$). The
candidates chosen from the list for the fibre-filling emphasised first
$z_{\rm phot} =$~0.7--0.9, 1.2--1.4, then $z_{\rm phot} =$~0.6--0.7,
0.9--1.2, 1.4--1.5, followed by any other $z_{\rm phot}$ values. However, the
later DR6 photometric quasar catalogue of \citet{Richards2009} revised the
photometric redshifts, sometimes substantially, and these criteria will have
become blurred. In retrospect, from the MMT / Hectospec spectroscopic
redshifts, the net effect seems to be that the fibre-filling selected
candidates typically with $z \sim$ 1.0--2.3 (90 per cent) and a few with $z
\sim$ 0.6--0.7 (10 per cent).

The MMT / Hectospec observations resulted in spectra for 31 quasars, with
redshifts in the range 0.613--2.228, and $g$ in the range 18.486--20.955. Of
these 31, 21 have $1.0 \le z \le 1.8$; these 21 have $g$ in the range
18.486--20.947. Note that we have taken magnitudes of the MMT / Hectospec
quasars from the later DR6 photometric quasar catalogue \citep{Richards2009}.
One of the quasars is seen to be a BAL quasar ($W2400 = 38.4$~\AA, strong).
Six of these 31 are also contained in the DR7QSO catalogue.

Accurate spectrophotometric calibration was not a particular concern of the
MMT / Hectospec observations and errors of $\la$~20 per cent are expected,
with the values of $L3000$ and hence log$_{10}$L3000
(Table~\ref{mmt_hecto_quasars}) affected correspondingly.

We note the consistency of the data in Table~\ref{mmt_hecto_quasars} with the
tendency of the strongest Fe~{\sc II} emitters to be associated with
relatively narrow Mg~{\sc II} emission. For $1.0 \le z \le 1.8$ and $sn\_med
\ge 10.0$, 9 of the 12 quasars (75 per cent) that are ultrastrong or strong
emitters have narrow $fwhm2798 < 35.0$~\AA, whereas only 2 of the 6 quasars
(33 per cent) that are weak emitters (and with Mg~{\sc II} unaffected by the
A-band) have $fwhm2798 < 35.0$~\AA.

We have also made a little use of ultraviolet spectra from GALEX. These
spectra were obtained in parallel with a main programme of imaging from GALEX
within the two MMT / Hectospec fields. The GALEX fields and the MMT /
Hectospec fields are essentially the same. The GALEX spectra are slitless, in
the far-UV (FUV), $\sim$~1350--1800~\AA, and the near-UV (NUV),
$\sim$~1800--2800~\AA. The point-source resolutions are $\sim$~10~\AA\ and
$\sim$~25~\AA\ respectively. Extraction of the spectra was done by GALEX
staff. GALEX procedures circumvent overlapped spectra by observing a field
with a series of grism-angle rotations relative to the sky\footnote{See
  www.galex.caltech.edu/DATA/gr1\_docs/grism/primer.html}. There can
be problems of second-order and third-order overlaps in the FUV affecting the
brighter and bluer objects. We have GALEX spectra for 11 of the MMT /
Hectospec quasars listed in Table~\ref{mmt_hecto_quasars}. Of these 11, five
provide usable spectra across the Ly$\alpha$ region. Two of these five
happened to have ultrastrong $W2400$; we used their spectra in
Fig.~\ref{w1216_galex_example_spectra}, together with the corresponding
optical spectra in Fig.~\ref{w2400_example_spectra}, to illustrate the
plausibility of broad / narrow Ly$\alpha$ corresponding to broad / narrow
Mg~{\sc II}. The other three have Ly$\alpha$ and Mg~{\sc II} widths that
correspond similarly.

\begin {table*}
\flushleft
\caption {A summary of our MMT / Hectospec quasar spectra. The columns are as
  follows. (1)~Category of the UV Fe~{\sc II} emission: ultrastrong ($W2400
  \ge 45$~\AA); strong ($30 \le W2400 < 45$~\AA); weak ($W2400 < 30$~\AA).
  (2)~Name of the quasar. A name beginning ``qso'' is our MMT / Hectospec
  name; if it is followed by a name beginning ``SDSSJ'' in parentheses then
  the same quasar is also in the SDSS DR7QSO catalogue \citep{Schneider2010}.
  (3)~Redshift $z$, obtained from the MMT / Hectospec spectra. (4)~RA,
  Dec. (2000).  (5)~Association of the quasars with the LQGs U1.11, U1.28,
  U1.54. The suffix ``o'' indicates that the quasar is an original member of
  the LQG (and so must have $i \le 19.1)$ from the DR7QSO catalogue. The
  suffix ``d'' indicates that the quasar is in the LQG domain, meaning that
  it is within the convex hull of member spheres (CHMS) of the original
  members.  See \citet{Clowes2012} and \citet{Clowes1991} for details of
  these LQGs; see \citet{Clowes2012} for the definition of CHMS.  (6)~Whether
  the quasar is known to be a BAL quasar. (7)~$W2400$ equivalent width (\AA)
  for the UV Fe~{\sc II} emission, as described in the text and in
  \citet{Clowes2013b}. (8)~$fwhm2798$ FWHM of the Mg~{\sc II} emission (\AA)
  as described in the text; if available it is given only if $W2400$ is also
  available.  (9)~$sn\_med$ s/n of the spectrum, as described in the
  text. (10,~11)~$g$, $i$ magnitudes, taken from the DR6 photometric quasar
  catalogue of \citet{Richards2009}.  (12)~$\log_{10}L3000$ with $L3000$ in
  units of erg~s$^{-1}$.
}
\scriptsize \renewcommand \arraystretch {0.8}
\renewcommand \tabcolsep{1.8pt}
\newdimen\padwidth
\setbox0=\hbox{\rm0}
\padwidth=0.3\wd0
\catcode`|=\active
\def|{\kern\padwidth}
\newdimen\digitwidth
\setbox0=\hbox{\rm0}
\digitwidth=0.70\wd0
\catcode`!=\active
\def!{\kern\digitwidth}
\begin {tabular} {llllllllllll}
\\
  (1)         & (2)                                & (3)    & (4)                         & (5)     & (6) & (7)   & (8)      & (9)     & (10)     & (11)    & (12)                        \\
  Category    & Quasar                             & z      & RA, Dec (2000)              & LQG     & BAL & W2400 & fwhm2798 & sn\_med & $g$      & $i$     & $\log_{10}L3000$            \\
              &                                    &        &                             &         &     & (\AA) & (\AA)    &         &          &         & ($\log_{10}$(erg~s$^{-1}$)) \\
\\
$1.0 \le z \le 1.8$ \\ \hline
  ultrastrong & qso412                             & 1.156  & 10:49:47.35  $+$04:17:46.3  & U1.11 d &     & 55.52 & 50.65    & 33.01   & 20.364   & 19.984  & 44.59                       \\
  ultrastrong & qso425 (SDSSJ104800.40$+$052209.7) & 1.230  & 10:48:00.41  $+$05:22:09.8  & U1.28 d &     & 52.49 & 24.22    & 52.79   & 19.607   & 19.129  & 44.97                       \\
  ultrastrong & qso417                             & 1.652  & 10:49:26.84  $+$04:23:34.7  &         &     & 50.71 & 20.81    & 25.15   & 20.086   & 19.649  & 44.87                       \\
  strong      & qso27                              & 1.313  & 10:49:30.45  $+$05:40:46.2  & U1.28 d &     & 44.40 & 26.46    & 13.46   & 20.947   & 20.805  & 44.21                       \\
  strong      & qso29                              & 1.416  & 10:49:21.06  $+$05:09:48.3  &         &     & 38.87 & 32.78    & 32.32   & 19.465   & 19.307  & 44.67                       \\
  strong      & qso421                             & 1.655  & 10:48:15.94  $+$05:50:07.8  &         & BAL & 38.39 & 26.67    & 17.20   & 20.487   & 20.396  & 44.70                       \\
  strong      & qso410                             & 1.421  & 10:50:00.37  $+$04:51:57.9  &         &     & 38.19 & 28.25    & 24.39   & 20.790   & 20.372  & 44.81                       \\
  strong      & qso48 (SDSSJ105010.05$+$043249.1)  & 1.215  & 10:50:10.06  $+$04:32:49.2  & U1.28 o &     & 37.19 & 24.38    & 70.32   & 18.486   & 18.087  & 45.36                       \\
  strong      & qso219                             & 1.348  & 10:49:34.71  $+$05:48:36.0  & U1.28 d &     & 33.09 & 25.04    & 23.10   & 20.797   & 20.408  & 44.50                       \\
  strong      & qso41                              & 1.430  & 10:51:31.95  $+$04:51:24.7  &         &     & 31.79 & 46.91    & 34.86   & 19.733   & 19.432  & 44.91                       \\
  strong      & qso416                             & 1.149  & 10:49:37.48  $+$04:57:57.1  & U1.11 d &     & 31.28 & 17.87    & 19.37   & 20.884   & 20.437  & 44.35                       \\
  strong      & qso24                              & 1.034  & 10:49:37.19  $+$05:45:19.2  &         &     & 30.12 & 40.71    & 30.13   & 20.297   & 20.032  & 44.45                       \\
  weak        & qso420$^a$                         & 1.237  & 10:48:40.85  $+$04:09:38.4  & U1.28 d &     & 29.27 & 25.21    & 08.95   & 20.325   & 19.664  & 44.32                       \\
  weak        & qso217                             & 1.619  & 10:49:58.91  $+$04:27:23.3  & U1.54 d &     & 28.93 & 25.20    & 25.45   & 20.700   & 20.393  & 44.82                       \\
  weak        & qso49 (SDSSJ105007.89$+$043659.7)  & 1.132  & 10:50:07.90  $+$04:36:59.8  & U1.11 d &     & 25.44 & 45.03    & 49.18   & 19.317   & 19.074  & 44.93                       \\
  weak        & qso22 (SDSSJ105030.75$+$043055.0)  & 1.215  & 10:50:30.76  $+$04:30:55.1  & U1.28 d &     & 24.21 & 27.63    & 69.27   & 19.742   & 19.214  & 45.06                       \\
  weak        & qso45                              & 1.311  & 10:50:36.09  $+$04:56:08.4  & U1.28 d &     & 23.27 & 46.73    & 31.47   & 20.856   & 19.917  & 44.73                       \\
  weak        & qso26 (SDSSJ104932.22$+$050531.7)  & 1.108  & 10:49:32.22  $+$05:05:31.7  & U1.11 o &     & 23.22 & 83.11    & 94.45   & 18.723   & 18.645  & 45.09                       \\
  weak        & qso225$^b$                         & 1.724  & 10:48:05.38  $+$05:39:37.3  &         &     & 19.11 & 67.84    & 19.19   & 20.704   & 20.361  &                             \\
  weak        & qso413 (SDSSJ104943.28$+$044948.8) & 1.295  & 10:49:43.29  $+$04:49:48.9  & U1.28 d &     & 17.54 & 56.47    & 63.83   & 19.437   & 19.079  & 45.25                       \\
  weak        & qso210$^b$                         & 1.738  & 10:49:14.94  $+$05:14:52.6  &         &     & 13.01 & 46.46    & 43.36   & 20.105   & 19.938  &                             \\
\\
Other $z$ \\ \hline
  ultrastrong & qso221                             & 1.832  & 10:49:18.73  $+$04:13:42.1  &         &     & 51.35 &          & 15.68   & 20.845   & 20.618  &                             \\
  weak        & qso223$^c$                         & 2.040  & 10:49:16.38  $+$05:48:26.0  &         &     & 22.35 &          & 21.61   & 20.306   & 20.061  &                             \\
  weak        & qso222$^c$                         & 1.951  & 10:49:18.14  $+$04:59:59.1  &         &     & 11.41 &          & 18.81   & 20.955   & 20.342  &                             \\
  weak        & qso215$^c$                         & 1.937  & 10:51:47.95  $+$04:43:12.0  &         &     & 00.76 &          & 36.67   & 20.512   & 20.157  &                             \\
\\
              & qso227                             & 0.663  & 10:47:58.05  $+$05:53:08.6  &         &     &       &          & 25.73   & 20.911   & 20.786  & 43.95                       \\
              & qso422                             & 2.129  & 10:48:10.71  $+$05:33:52.8  &         &     &       &          &         & 19.691   & 19.660  &                             \\
              & qso28                              & 0.613  & 10:49:26.39  $+$05:09:02.5  &         &     &       &          & 26.52   & 20.170   & 19.976  & 43.85                       \\
              & qso25                              & 2.109  & 10:49:32.87  $+$05:19:28.2  &         &     &       &          &         & 20.946   & 20.907  &                             \\
              & qso218                             & 2.228  & 10:49:42.18  $+$04:48:14.0  &         &     &       &          &         & 20.790   & 20.743  &                             \\
              & qso415                             & 0.696  & 10:49:42.39  $+$04:18:23.4  &         &     &       &          & 35.77   & 19.347   & 19.215  & 44.11                       \\

\end {tabular}
~\\
~\\
~\\
$^a$ Fails to meet our criterion $sn\_med \ge 10.0$ for reliable measurements.                                            \\
$^b$ The Mg~{\sc II} emission contains the interpolation across the atmospheric A-band.                                   \\
$^c$ The Fe~{\sc II} emission contains the interpolation across the atmospheric A-band.                                   \\
$W2400$ and $fwhm2798$ have been retained with ``spurious precision'' to avoid tied values if used in statistical tests.  \\
\label{mmt_hecto_quasars}
\end {table*}

\bsp

\label{lastpage}

\end{document}